\documentclass[aps,prx,reprint,superscriptaddress,longbibliography,nofootinbib]{revtex4-2}
\usepackage[T1]{fontenc}
\usepackage{lmodern}
\usepackage{amsmath,amssymb,amsthm,bm}
\usepackage{graphicx}
\graphicspath{{../figures/}}
\usepackage[dvipsnames]{xcolor}
\usepackage{booktabs}
\usepackage[colorlinks=true,linkcolor=blue,citecolor=blue,urlcolor=blue]{hyperref}
\usepackage{orcidlink}
\usepackage{tikz}
\usetikzlibrary{positioning,fit,backgrounds,arrows.meta,calc,decorations.pathmorphing,decorations.pathreplacing,shadows.blur}
\definecolor{fBlue}{HTML}{2171B5}
\definecolor{fRed}{HTML}{CB181D}
\definecolor{fDark}{HTML}{08306B}
\definecolor{fLight}{HTML}{C6DBEF}

\newtheorem{lemma}{Lemma}
\newtheorem{proposition}{Proposition}
\newtheorem{corollary}{Corollary}

\newcommand{\SEEAPPENDIX}[1]{#1}
\definecolor{OverlapBlue}{HTML}{00008B}
\newcommand{\CHANGE}[1]{\textcolor{black}{#1}}

\newcommand{\NL}{\mathrm{NL}}
\newcommand{\F}{\mathcal{F}}

\newcommand{\Sf}{\mathcal{S}_{\flat}}
\newcommand{\Dm}{D^{f,\NL}_{\min}}
\newcommand{\Qsh}{Q_{\rm sh}}
\providecommand{\Tr}{\operatorname{Tr}}
\providecommand{\ket}[1]{|#1\rangle}
\providecommand{\bra}[1]{\langle#1|}

\providecommand{\braket}[2]{\langle#1|#2\rangle}
\newcommand{\SMref}[1]{Sec.~\ref{#1} of the SM}

\begin{document}

\title{Nonlocal magic of fermions is bounded by entanglement fluctuations and is logarithmic for free fermions}

\author{Raghav~G.~Jha\orcidlink{0000-0003-2933-0102}}
\email{raghav.govind.jha@gmail.com}
\affiliation{Department of Physics and Astronomy, North Carolina State University, Raleigh, North Carolina 27695, USA}

\author{Dengyao Luo\orcidlink{0009-0004-2383-403X}}
\email{dluo6@ncsu.edu}
\affiliation{Department of Mathematics, North Carolina State University, Raleigh, North Carolina 27695, USA}

\author{Alexander~F.~Kemper\orcidlink{0000-0002-5426-5181}}
\affiliation{Department of Physics and Astronomy, North Carolina State University, Raleigh, North Carolina 27695, USA}

\author{Bojko~N.~Bakalov\orcidlink{0000-0003-4630-6120}}
\affiliation{Department of Mathematics, North Carolina State University, Raleigh, North Carolina 27695, USA}

\date{\today}

\begin{abstract}
Free-fermion states are classically simulable, yet, written as qubits, they typically carry extensive magic, the quantum resource that stabilizer circuits lack. Part of this magic is nonlocal: local unitaries acting separately on two parts of the system cannot remove it. If these local unitaries are restricted to Gaussian ones, the nonlocal magic grows \emph{linearly} with the number $m$ of partially entangled fermion pairs. We show that this growth is an artifact of the Gaussian restriction: local Gaussian unitaries minimize the nonlocal stabilizer R\'enyi entropy for up to three pairs, but not from four pairs on. General local non-Gaussian unitaries that respect fermion parity take $m$ free-fermion pairs to fermionic Bell pairs with a fidelity that decays at most as $1/\sqrt m$ compared to exponential decay with Gaussian unitaries, so the nonlocal magic drops from linear to logarithmic in $m$. The reason is \emph{statistical} --- Bell pairs have equal Schmidt weights, that is, a fully degenerate entanglement spectrum, and a state can stay far from them only if its Schmidt weights are spread over many orders of magnitude. This spread is measured by the capacity of entanglement $C_E$, the variance of the entanglement Hamiltonian, and the nonlocal magic of \emph{any} fermionic state grows at most logarithmically with $C_E$.
Free-fermion pairs add to $C_E$ independently, so $C_E$ grows only linearly in $m$, and this leads to the stabilizer 
fidelity decreasing at most as $1/\sqrt m$. Almost all the magic of free fermions, including Hartree--Fock and BCS mean-field states, is therefore \emph{local}. 
\end{abstract}

\maketitle

\section{Introduction}\label{sec:m-intro}

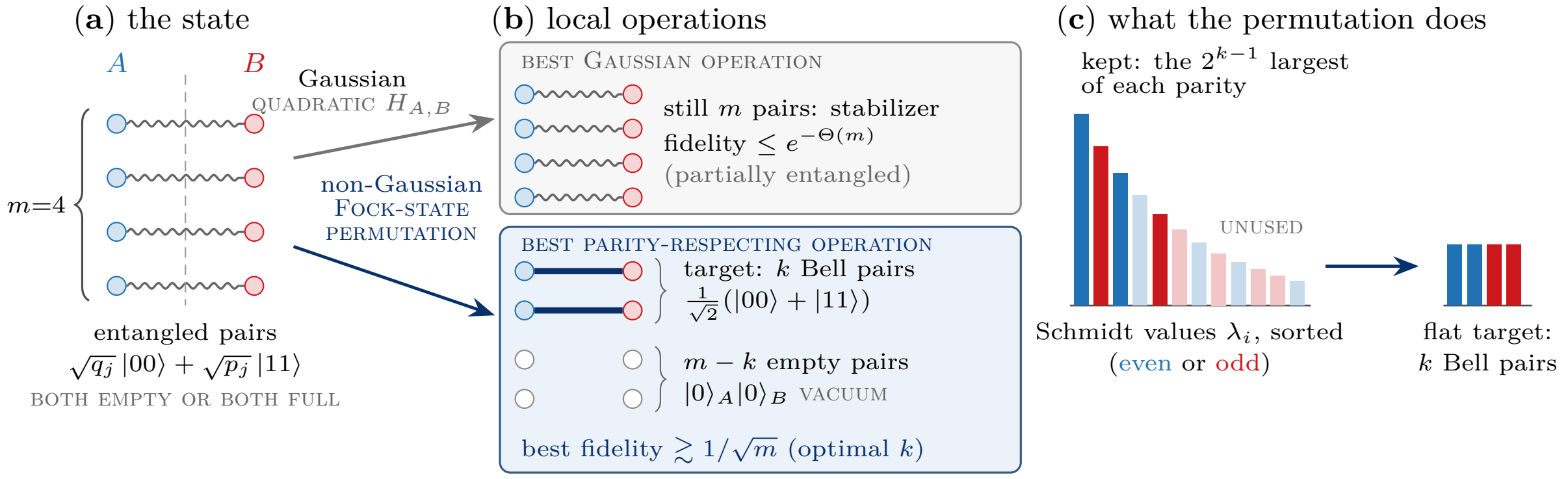
\begin{figure*}[t]
\centering
\begin{tikzpicture}[font=\footnotesize, x=1cm, y=1cm,
  mA/.style={circle, draw=fBlue, fill=fBlue!20, inner sep=1.9pt},
  mB/.style={circle, draw=fRed, fill=fRed!18, inner sep=1.9pt},
  mE/.style={circle, draw=black!45, fill=white, inner sep=1.9pt},
  bcs/.style={black!60, line width=0.65pt, decorate, decoration={snake, amplitude=0.8pt, segment length=4pt}},
  bell/.style={fDark, line width=1.8pt},
  arr/.style={-{Stealth[length=2.6mm,width=2mm]}, line width=0.9pt},
  ttl/.style={font=\small, anchor=west, inner sep=0pt}]
\node[ttl] at (-0.25,0.55) {(\textbf{a}) the state};
\foreach \i in {1,...,4} {
  \node[mA] (a\i) at (0.2, -0.55*\i+0.05) {};
  \node[mB] (b\i) at (1.6, -0.55*\i+0.05) {};
  \draw[bcs] (a\i) -- (b\i);
}
\draw[densely dashed, black!40] (0.9,0.0) -- (0.9,-2.35);
\node[fBlue] at (0.2,0.12) {$A$}; \node[fRed] at (1.6,0.12) {$B$};
\draw[decorate, decoration={brace, amplitude=4pt}, black!70, line width=0.6pt] (-0.08,-2.3) -- (-0.08,-0.35);
\node[anchor=east, font=\scriptsize] at (-0.2,-1.325) {$m{=}4$};
\node[align=center, font=\scriptsize] at (0.9,-2.95) {entangled pairs\\[1pt]$\sqrt{q_j}\,\ket{00}+\sqrt{p_j}\,\ket{11}$\\[2pt]{\color{black!60}\textsc{both empty or both full}}};
\node[ttl] at (4,0.55) {(\textbf{b}) local operations};
\draw[arr, black!55] (2.0,-0.85) -- (4.05,-0.45);
\node[align=center, font=\scriptsize, anchor=south] at (2.6,-0.55) {Gaussian\\[-1pt]{\color{black!60}\textsc{quadratic} $H_{A,B}$}};
\foreach \i in {1,...,4} {
  \node[mA] (g\i) at (4.35, -0.35*\i+0.15) {};
  \node[mB] (h\i) at (5.45, -0.35*\i+0.15) {};
  \draw[bcs] (g\i) -- (h\i);
}
\node[anchor=west, align=left, font=\scriptsize] at (5.65,-0.7) {still $m$ pairs: stabilizer\\[1pt]fidelity $\le e^{-\Theta(m)}$\\[1pt]{\color{black!60}(partially entangled)}};
\draw[arr, fDark] (2.0,-1.75) -- (4.05,-2.45);
\node[align=center, font=\scriptsize, anchor=south, fDark] at (3.1,-1.8) {non-Gaussian\\[-1pt]\textsc{Fock-state}\\[-1pt]\textsc{permutation}};
\node[mA] (c1) at (4.35,-2.0) {}; \node[mB] (d1) at (5.45,-2.0) {};
\node[mA] (c2) at (4.35,-2.4) {}; \node[mB] (d2) at (5.45,-2.4) {};
\node[mE] at (4.35,-2.9) {}; \node[mE] at (5.45,-2.9) {};
\node[mE] at (4.35,-3.3) {}; \node[mE] at (5.45,-3.3) {};
\draw[bell] (c1) -- (d1); \draw[bell] (c2) -- (d2);
\draw[decorate, decoration={brace, amplitude=3pt, mirror}, black!60] (5.68,-2.53) -- (5.68,-1.87);
\draw[decorate, decoration={brace, amplitude=3pt, mirror}, black!60] (5.68,-3.43) -- (5.68,-2.77);
\node[anchor=west, align=left, font=\scriptsize] at (5.85,-2.2) {target: $k$ Bell pairs\\[1pt]$\tfrac1{\sqrt2}(\ket{00}+\ket{11})$};
\node[anchor=west, align=left, font=\scriptsize] at (5.85,-3.1) {$m-k$ empty pairs\\[1pt]$\ket0_A\ket0_B$ {\color{black!60}\textsc{vacuum}}};
\node[anchor=west, font=\scriptsize, fDark] at (4.2,-3.82) {best fidelity $\gtrsim1/\sqrt m$ (optimal $k$)};
\begin{scope}[on background layer]
  \draw[rounded corners=3pt, black!45, fill=black!3, line width=0.6pt] (4.1,0.33) rectangle (9.4,-1.42);
  \draw[rounded corners=3pt, fDark!80, fill=fLight!35, line width=0.6pt] (4.1,-1.55) rectangle (9.4,-4.05);
\end{scope}
\node[anchor=west, font=\scriptsize, text=black!65] at (4.2,0.15) {\textsc{best Gaussian operation}};
\node[anchor=west, font=\scriptsize, text=fDark] at (4.2,-1.73) {\textsc{best parity-respecting operation}};
\node[ttl] at (9.75,0.55) {(\textbf{c}) what the permutation does};
\begin{scope}[shift={(9.9,-2.35)}]
  \draw[black!70, line width=0.5pt] (0,0) -- (2.45,0);
  \foreach \h/\c [count=\n from 0] in {1.95/fBlue, 1.62/fRed, 1.35/fBlue, 1.12/fBlue!25, 0.93/fRed, 0.77/fRed!25, 0.64/fBlue!25, 0.53/fRed!25, 0.44/fBlue!25, 0.37/fRed!25, 0.30/fRed!25, 0.25/fBlue!25}
    { \fill[\c] (0.04+0.2*\n,0) rectangle ++(0.15,\h); }
  \node[anchor=north, align=center, font=\scriptsize] at (1.22,-0.06) {Schmidt values $\lambda_i$, sorted\\[1pt]({\color{fBlue}even} or {\color{fRed}odd})};
  \node[anchor=south west, align=left, font=\scriptsize] at (0.0,2.0) {kept: the $2^{k-1}$ largest\\[-1pt]of each parity};
  \node[font=\scriptsize, black!55, align=center] at (1.95,0.8) {\textsc{unused}};
  \draw[arr, fDark] (2.6,0.4) -- (3.55,0.4);
  \draw[black!70, line width=0.5pt] (3.8,0) -- (4.7,0);
  \foreach \c [count=\n from 0] in {fBlue, fBlue, fRed, fRed} { \fill[\c] (3.84+0.2*\n,0) rectangle ++(0.15,0.62); }
  \node[anchor=north, align=center, font=\scriptsize] at (4.25,-0.06) {flat target:\\[1pt]$k$ Bell pairs};
\end{scope}
\end{tikzpicture}
\caption{Free-fermion entanglement conversion into Bell pairs. In \CHANGE{panels (a,b)} we take $m=4, k=2$\CHANGE{; panel (c) is schematic}. (a) Across a cut, a free-fermion (Gaussian) state is, in suitably chosen modes, a product of $m$ entangled (BCS) pairs $\sqrt{q_j}\ket{00}+\sqrt{p_j}\ket{11}$~\cite{Botero2004BCS}. (b) The two parties apply local parity-respecting unitaries, without measuring or communicating, and aim at Bell pairs, the stabilizer states shared by two parties up to local unitaries. Best Gaussian operation (top box): for partially entangled pairs, with occupations bounded away from $0$, $\frac12$, and $1$, the fidelity with every stabilizer state is exponentially small in $m$. Best operation overall (bottom box): after the Gaussian rotation to BCS form, a Fock-state permutation (non-Gaussian) reaches $k$ Bell pairs plus empty modes, 
with fidelity at least of order $1/\sqrt m$ (Sec.~\ref{sec:m-free}). (c) The permutation places the $2^{k-1}$ largest Schmidt values of each parity (blue: even, red: odd) on the target. Parity forbids mixing the colors, so the fourth-largest value, which is even, is skipped in favor of a smaller odd one; this is the \emph{cost} of parity superselection. The fidelity is $\Phi_k=2^{-k}(\sum_{\rm kept}\sqrt{\lambda_i})^2$, maximized over $k$ [Eq.~\eqref{eq:m-PhiS}].}\label{fig:m-scheme}
\end{figure*}

What makes a quantum computation hard to simulate classically, and what makes it costly on a fault-tolerant machine? For qubits, both questions lead to magic, or nonstabilizerness. Clifford circuits acting on stabilizer states are simulated efficiently with the stabilizer formalism~\cite{Gottesman:1998hu}. 
In addition, magic is also the expensive ingredient of fault-tolerant quantum computation. Due to the Eastin--Knill theorem, no quantum error-correcting code admits a universal set of transversal logical gates~\cite{EastinKnill2009}, and fault-tolerant architectures usually implement the missing non-Clifford gates by consuming magic states, which are prepared in dedicated distillation factories~\cite{BravyiKitaev2005,Litinski2019}. Magic thus measures both the cost of classical simulation by stabilizer methods and the cost of fault-tolerant execution. Free-fermion (matchgate) circuits are classically easy to simulate by different methods~\cite{JozsaMiyake2008,BravyiKitaev2002,kokcu2022algebraic,camps2022algebraic}, even though their states generally carry magic. The two classes (matchgates and stabilizer circuits) become universal when combined~\cite{JozsaMiyake2008,BrodGalvao2011,Projansky2024} leading to the conclusion that:
\emph{computational/circuit classes that are easy separately can be hard together}. Fermionic states are also among the main targets of quantum computation. Free-fermion (Gaussian), a special class of fermionic states, include the Hartree--Fock and BCS (mean-field) states that serve as reference states in quantum chemistry and condensed matter, as initial states of quantum algorithms~\cite{Kivlichan2018,hogan2026efficient}, as building blocks of classical methods for interacting fermions~\cite{Fishman2015,Krumnow2016,BravyiGosset2017}, and as the native states of fermionic quantum processors~\cite{BravyiKitaev2002,GonzalezCuadra2023}\CHANGE{, while the correlated ground states of interacting models are what such algorithms aim to prepare. Where the magic of fermionic states resides, and how much of it must be carried by the entanglement between two parts of a system, is the question we address here.}

\CHANGE{Consider a pure fermionic state shared by two regions $A$ and $B$. Part of its magic is local, since unitaries acting on $A$ and $B$ separately can remove it. What remains is the nonlocal magic, the magic stored in the entanglement between the regions~\cite{Bao2021NonlocalMagic,Cao2025Gravitational}. A two-party stabilizer state is a collection of Bell pairs and unentangled qubits once each party has chosen a suitable basis, so a state has no nonlocal magic exactly when local unitaries can turn it into such a state. Nonlocal magic therefore measures how far the entanglement of the state is from that of Bell pairs, and it depends only on the Schmidt spectrum across the cut (for fermions, together with the parities of the Schmidt vectors). It is also the part of the magic that unitary operations within the regions cannot remove: it bounds from below the number of non-Clifford ($T$) gates, and hence of magic states, needed to prepare the state up to local unitaries. For fermions the local unitaries must commute with the fermion parity of each region (parity superselection), and we allow every such unitary, Gaussian or not. Our general results hold for every pure fermionic state of definite parity, \emph{free or interacting}: an exact result for the largest fidelity with a stabilizer state that parity-even local unitaries can reach, the statement that parity superselection costs at most one bit, and a bound on the nonlocal magic by the capacity of entanglement $C_E$, the variance of the entanglement Hamiltonian.}

\CHANGE{Written as qubits, free-fermion states typically carry extensive magic~\cite{Collura2026Gaussian,Turkeshi2026RandomGaussian}. Across any cut, they are, up to local Gaussian unitaries, a product of $m$ entangled (BCS) pairs: one mode in $A$ and one in $B$ that are both empty or both occupied [Fig.~\ref{fig:m-scheme}(a)]. How much of their magic is nonlocal depends on which local operations are allowed. If the local unitaries are restricted to Gaussian ones, \CHANGE{which map free-fermion states to free-fermion states,} the nonlocal magic grows linearly with the number $m$ of partially entangled pairs~\cite{Iannotti2026FNL,Collura2026NLFF}. General local unitaries may be non-Gaussian, but for fermions they are still constrained by superselection: they must commute with the fermion parity of each region [Fig.~\ref{fig:m-scheme}(b)].}

\CHANGE{Non-Gaussian local unitaries can only lower the Gaussian-restricted value, and for subsystems of at most two modes it was shown that they cannot~\cite{Iannotti2026FNL}. General bounds for qubits show that the nonlocal magic of any state is at most logarithmic in the system size~\cite{Sierant2026Exact}, and Aditya, Sierant, and Turkeshi found that for free fermions it is logarithmic in the Gaussian-restricted value under mild conditions~\cite{Aditya2026Spreading} (Sec.~\ref{sec:m-background}). Two questions remain: (1) what the nonlocal magic of free fermions actually is, rather than how large it can be, and (2) what changes when the local operations must respect parity superselection but may be non-Gaussian. In this paper, we answer these and some further questions. Our main results are as follows.}
\begin{itemize}
\item[(i)] \CHANGE{Parity-even local unitaries including non-Gaussian unitaries i.e., a Gaussian mode rotation followed by a permutation of Fock states [Fig.~\ref{fig:m-scheme}(b,c)], bring $m$ partially entangled pairs to fermionic Bell pairs with fidelity of order $1/\sqrt m$, an order that cannot be improved. Due to this, both measures of magic (Sec.~\ref{sec:m-background}) drop from $\Theta(m)$ to $\Theta(\log m)$\footnote{$f(m)=\Theta(g(m))$ means $c_1g(m)\le f(m)\le c_2g(m)$ for large $m$, with constants $c_1,c_2>0$.} when the local unitaries are allowed to be parity-preserving non-Gaussian unitaries. For up to three pairs, however, local Gaussian unitaries already minimize the SRE and the distinction between Gaussian and non-Gaussian parity preserving unitaries is not essential.}

\item[(ii)] \CHANGE{The slow $1/\sqrt m$ decay of fidelity with stabilizer state has a statistical origin. We read the Schmidt weights $\lambda_i$ as probabilities of the entanglement energies $X_i=-\log_2\lambda_i$, measured in bits.
Bell pairs have a flat spectrum, with all $X_i$ equal, so the best Bell-pair fidelity is essentially the probability that $X$ falls into one window. This probability can be small only if $X$ is spread over many bits, that is, only if its variance $C_E/\ln^22$ is large, where $C_E$ is the capacity of entanglement. This gives a bound for every state (free or interacting): every pure fermionic state of definite parity can be brought by parity-even local unitaries to a stabilizer state with fidelity at least $0.478/\sqrt{C_E+5}$. For free fermions, $X$ is a sum of independent contributions, one per pair. Each pair adds at most $c_\ast\approx0.44$ to $C_E$, so $X$ spreads over only about $\sqrt m$ bits and the fidelity decays only as $1/\sqrt m$ (Sec.~\ref{sec:m-free}).}
\item[(iii)] \CHANGE{When the distribution of $X$ becomes Gaussian as $C_E\to\infty$, the nonlocal magic follows capacity laws: $\frac12\log_2C_E$ for the fidelity-based measure and $\frac32\log_2C_E$ for the SRE. For a critical block of $\ell$ sites it \CHANGE{should then grow} only as $\log\log\ell$ (Sec.~\ref{sec:m-laws-par}).}
\end{itemize}
\CHANGE{The plan of the paper is as follows. Section~\ref{sec:m-background} defines the measures of magic and collects the results on which we build. Sections~\ref{sec:m-exact}--\ref{sec:m-free} develop our results, and Sec.~\ref{sec:discussion} discusses their consequences and limitations. The Supplemental Material (SM) contains the proofs and further remarks.}

\section{Measures of nonlocal magic and prior results}\label{sec:m-background}

In this section, we define the measures of magic used in this paper and collect 
the results on which we build. In the context of stabilizer states,
the magic of a state $\ket\psi$ of $N$ qubits is quantified by the stabilizer R\'enyi entropy (SRE)~\cite{Leone2022SRE},
\begin{equation}
M_\alpha(\ket\psi)=\frac1{1-\alpha}\log_2\Big(2^{-N}\sum_P\bra\psi P\ket\psi^{2\alpha}\Big), \label{eq:m-SRE}
\end{equation}
typically using $\alpha=2$, 
with the sum over all $4^N$ Pauli strings $P$. $M_\alpha(\psi)$ vanishes if and only if $\ket\psi$ is a stabilizer state. For a state shared by two regions $A$ and $B$, part of the magic is local: unitaries acting on $A$ and $B$ separately can remove it. What remains
is the nonlocal magic~\cite{Bao2021NonlocalMagic,Cao2025Gravitational},
\begin{equation}
M^{\NL}_\alpha(\ket\psi)=\min_{U_A\otimes U_B}M_\alpha\big(U_A\otimes U_B\ket\psi\big), \label{eq:m-MNL}
\end{equation}
 i.e., the magic stored in the correlations between $A$ and $B$. \CHANGE{The minimum runs over all local unitaries, so $M^{\NL}_\alpha$ is a property of $\ket\psi$ alone; for a free-fermion state $\ket\psi$ the minimizing unitaries need not be Gaussian.}
\CHANGE{A second measure of nonlocal magic is built from fidelities instead of Pauli moments. It is often easier to compute, since for qubits it is given by an explicit formula in the entanglement spectrum (Sec.~\ref{sec:m-exact}), and it has a direct operational meaning. Let ${\rm STAB}$ denote the set of pure stabilizer states of the $N$ qubits. The largest fidelity of $\ket\psi$ with a stabilizer state defines the min-relative entropy of magic, $D_{\min}(\ket\psi)=-\log_2\max_{\ket s\in{\rm STAB}}|\braket s\psi|^2$, which, like $M_\alpha$, vanishes exactly on stabilizer states. Maximizing in addition over local unitaries gives the nonlocal stabilizer fidelity $\Phi^{\rm q}$ and the nonlocal min-relative entropy of magic (NMRE) $D^{\NL}_{\min}$~\cite{Sierant2026Exact},
\begin{equation}
\begin{gathered}
\Phi^{\rm q}(\ket\psi)=\max_{U_A\otimes U_B}\ \max_{\ket s\in{\rm STAB}}\big|\bra s\,U_A\otimes U_B\ket\psi\big|^2,\\
D^{\NL}_{\min}(\ket\psi)=-\log_2\Phi^{\rm q}(\ket\psi),
\end{gathered} \label{eq:m-Phiq}
\end{equation}
where the superscript ${\rm q}$ indicates that $U_A$ and $U_B$ are arbitrary unitaries on the qubits of $A$ and $B$.} Since a two-party stabilizer state is, up to local unitaries, a collection of Bell pairs and unentangled qubits, $\Phi^{\rm q}$ measures how well $A$ and $B$ can turn their state into Bell pairs by local unitaries alone, without measuring or communicating. For qubits it depends only on the entanglement spectrum (ES)~\cite{Sierant2026Exact,Viscardi2026Purification}, and it is at most logarithmic for every state: $D^{\NL}_{\min}\le\log_2(\nu+1)$ and $D^{\NL}_{\min}\le\log_2(2S_1+3)$, with $\nu$ the number of qubits of the smaller side and $S_1$ the entanglement entropy in bits, and $M^{\NL}_2\le4D^{\NL}_{\min}$~\cite{Sierant2026Exact}; see also Refs.~\cite{Sierant2026Spectral,Liu2026AntiFlatness,Torre2026NonLocal,Aditya2026Spreading}. For fermions, parity superselection restricts $U_A$ and $U_B$ in Eq.~\eqref{eq:m-MNL} to parity-even unitaries, which defines the fermionic nonlocal SRE $M^{f,\NL}_\alpha$ and its fidelity-based counterpart $\Dm$ (Sec.~\ref{sec:m-exact}).

Gaussian states are fixed by their two-point correlations and evolve efficiently under quadratic Hamiltonians, even at arbitrarily long times~\cite{BravyiKitaev2002,Hebenstreit2019}. Written as qubits, however, they typically carry extensive magic~\cite{Collura2026Gaussian,Turkeshi2026RandomGaussian}. This does not contradict their simulability, because magic measures the cost of a state for the stabilizer formalism, not classical hardness in general; it does force a Clifford$+T$ preparation to use a number of $T$ gates at least proportional to $N$~\cite{Zhu2024Amortized}, although polynomially many suffice to prepare it to any fixed accuracy~\CHANGE{\cite{Jiang2018Gaussian,RossSelinger2016}}. Across any cut, a pure Gaussian state is, up to local Gaussian unitaries, a product of $m$ BCS pairs~\footnote{For any pure Gaussian state and any cut, a Gaussian unitary on $A$ and one on $B$ bring the state to the fermionic Schmidt (BCS) normal form of Botero and Reznik~\cite{Botero2004BCS}, $\bigotimes_{j=1}^{m}\big(\sqrt{q_j}\,\ket{00}+\sqrt{p_j}\,\ket{11}\big)\otimes(\text{unpaired modes, each empty or filled})$.} $\sqrt{q_j}\ket{00}+\sqrt{p_j}\ket{11}$, each made of one mode in $A$ and one in $B$ that are both empty or both occupied, with occupation $p_j$ and $q_j=1-p_j$ [Fig.~\ref{fig:m-scheme}(a)]. Minimizing the SRE over local Gaussian unitaries (denoted $\mathfrak{M}$ to differentiate from $M^{\NL}_2$) alone gives~\cite{Iannotti2026FNL,Collura2026NLFF} \SEEAPPENDIX{(\SMref{sec:setting})}, for every pure Gaussian state [Theorem~1 of Ref.~\cite{Iannotti2026FNL},
\begin{equation}
\mathfrak M_2(p)=-\sum_j\log_2\big[1-4p_jq_j(1-2p_j)^2\big]. \label{eq:m-Mfree2}
\end{equation}
With $p_j=\sin^2\theta_j$, each term is $-\log_2(1-\frac14\sin^24\theta_j)$. It vanishes only for $p_j\in\{0,\frac12,1\}$, where the pair is a product state or a Bell pair, and it is largest, $\log_2\frac43$, at $\theta_j=\pi/8$ \CHANGE{(and $3\pi/8$)}. We call the pairs \emph{partially entangled} if their occupations stay bounded away from these values, $p_j\in[\delta,\frac12-\delta]\cup[\frac12+\delta,1-\delta]$ for a fixed $\delta>0$. For such partially entangled pairs the Gaussian-restricted nonlocal magic is extensive, $\mathfrak M_2=\Theta(m)$. 

\paragraph*{The capacity of entanglement.} One number bounds the nonlocal magic in this paper. Write the state in Schmidt form, $\ket\psi=\sum_i\sqrt{\lambda_i}\,\ket{u_i}_A\ket{v_i}_B$; the Schmidt values $\lambda_i$, the ES, are the eigenvalues of the reduced density matrix $\rho_A$. Writing $\rho_A=e^{-H_E}$ defines the entanglement Hamiltonian $H_E=-\ln\rho_A$~\cite{LiHaldane2008}: $\rho_A$ is a thermal (Gibbs) state of $H_E$ at unit temperature, whose energy levels, the entanglement energies \CHANGE{$-\ln\lambda_i=X_i\ln2$, with $X_i=-\log_2\lambda_i$ the same energies in bits,} are occupied with probabilities $\lambda_i$. The mean energy is the entanglement entropy, in natural units, and the variance of the entanglement energy is the capacity of entanglement~\cite{Yao2010Capacity,deBoer2019Capacity},
\begin{equation}
\begin{aligned}
C_E&={\rm Var}\,H_E=\ln^2\!2\;{\rm Var}\,X\\
&=\sum_i\lambda_i\ln^2\lambda_i-\Big(\sum_i\lambda_i\ln\lambda_i\Big)^2,
\end{aligned}\label{eq:m-CE}
\end{equation}
the heat capacity of $H_E$. It measures the spread of the entanglement energies, not their number: it vanishes for a flat ES, such as that of Bell pairs or of a product state, however large the entanglement entropy, and it is large only when Schmidt values that differ by many orders of magnitude all carry appreciable weight; a dominant value with a faint tail, $\lambda=(0.999,0.001)$, has $C_E\approx0.05$. For one BCS pair, whose Schmidt values are $q$ and $p$, it is $pq\ln^2(q/p)$, which vanishes for a product pair and for a Bell pair; for independent pairs the capacities add (Sec.~\ref{sec:m-bcs}).

\paragraph*{Relation to prior work.} For qubits, the exact nonlocal stabilizer fidelity is due to Refs.~\cite{Sierant2026Exact,Viscardi2026Purification}, the bound result $M_2\le4D_{\min}$ is due to Ref.~\cite{Haug2023Stabilizer}, and its nonlocal form and the universal bounds above to Ref.~\cite{Sierant2026Exact}, the dyadic-shell characterization of the nonlocal SRE to Ref.~\cite{Sierant2026Spectral}, and the doubly logarithmic growth of nonlocal magic at criticality to Refs.~\cite{Sierant2026Spectral,Liu2026AntiFlatness,Torre2026NonLocal}, with the slope $\frac32$ of the SRE derived for the ordered reference state in Ref.~\cite{Liu2026AntiFlatness} and, via the dyadic-shell theorem, for the Ising chain in Ref.~\cite{Sierant2026Spectral}. Reference~\cite{Aditya2026Spreading} connected nonlocal magic to $C_E$, bounded it by $\frac12\log_2(1+C_E/\ln^22)+\log_2\frac{160}3$, derived the capacity law for spectra that are smooth on the scale of one bit, and, through its Eq.~(B3), implies the order $\frac12\log_2m$ for partially entangled free-fermion pairs of qubits.

In this paper, the new results are as follows --- (a) the treatment of parity superselection: the exact fidelity, its cost of at most one bit, and the contrast with particle-number superselection; (b) an exact kernel identity and, from it, a capacity bound that, like that of Ref.~\cite{Aditya2026Spreading}, holds at every size, but is sharper; (c) explicit constants at every size for the free-fermion order $1/\sqrt{m}$, including superselection, against $1/m$ for general states; (d) the capacity laws which we expect critical hopping chains to satisfy, with a lower bound on the asymptotic offset of the SRE; and (e) the sharp crossover at small sizes: extending the two-mode statement of Ref.~\cite{Iannotti2026FNL} \CHANGE{by a proof that} local Gaussian unitaries are optimal among parity-even ones for up to \emph{three pairs}, and from four pairs on they are not always optimal [Fig.~\ref{fig:m-laws}(d)] \SEEAPPENDIX{(\SMref{sec:small})}.

\section{Bell pairs from fermionic entanglement}\label{sec:m-exact}

\subsection{\texorpdfstring{\CHANGE{The exact fidelity under superselection}}{The exact fidelity under superselection}}\label{sec:m-exactfid}

In the Schmidt form $\ket\psi=\sum_i\sqrt{\lambda_i}\,\ket{u_i}_A\ket{v_i}_B$ of Sec.~\ref{sec:m-background}, local unitaries rotate the Schmidt vectors but leave the ES $\lambda$ unchanged, and any two states with the same ES are related by local unitaries, so every nonlocal quantity is a function of $\lambda$. In particular, the largest fidelity that local unitaries can reach between two states with spectra $\lambda$ and $\mu$ is $\big(\sum_i\sqrt{\lambda_i\mu_i}\big)^2$, with both spectra sorted in decreasing order. The ES of $k$ Bell pairs is flat, $2^k$ values equal to $2^{-k}$. For qubits, setting $\mu_i=2^{-k}$ for $i\le2^k$ gives the fidelity of the best conversion into Bell pairs~\cite{Sierant2026Exact,Viscardi2026Purification}, $\Phi^{\rm q}=\max_k2^{-k}\big(\sum_{i\le2^k}\sqrt{\lambda_i}\big)^2$: one lays a flat template of $2^k$ equal amplitudes over the $2^k$ largest Schmidt values and picks the best $k$. 

Fermions obey parity superselection first established by Wick, Wightman, and Wigner (WWW)~\cite{WWW1952} --- physical operations on $A$ must commute with its fermion parity $P_A$, so local unitaries must be \emph{parity-even}.\footnote{\CHANGE{WWW showed that no observable connects states of even and odd fermion number (in their terms, integer and half-integer angular momentum). A Hamiltonian acting on $A$ or $B$ alone is therefore even in the Majorana operators of that region, and so is every unitary it generates.}} For a state of definite total parity the Schmidt vectors can be chosen with definite parity, and the ES splits into an even and an odd sector, $\lambda=(\lambda^{(+)},\lambda^{(-)})$, the \emph{graded} ES. Parity-even unitaries cannot move a Schmidt value from one sector to the other, and two states of equal total parity are related by them if and only if their graded spectra coincide \SEEAPPENDIX{(\SMref{sec:exact})}.

Parity also restricts the flat spectra that are available. Only stabilizer states of definite parity matter as targets: projecting a stabilizer state with $\langle P\rangle=0$ onto the parity of $\ket\psi$ gives another stabilizer state with twice the fidelity \SEEAPPENDIX{(\SMref{app:exact})}. These are the Majorana stabilizer states, and neither they nor $M_\alpha$ depend on the Jordan--Wigner ordering, since reordering modes takes fermionic swaps, which are Clifford gates. For a stabilizer state of definite parity, either $\pm P_A$ is itself a stabilizer and all $2^k$ levels lie in one sector (a \emph{pure} split), or $\langle P_A\rangle=0$ and the levels are shared equally by the two sectors (a \emph{balanced} split). A balanced split is the ES of $k$ fermionic Bell pairs $(\ket{00}+\ket{11})/\sqrt2$; a pure split is that of $k$ Bell pairs together with one mode on each side that stores their total parity. Unequal splits, such as three even levels and one odd, which qubits allow, are excluded. Our figure of merit is the nonlocal stabilizer fidelity
\begin{equation}
\Phi^{\NL}_S=\max_{\substack{U_A\otimes U_B\\ [U_A,P_A]=[U_B,P_B]=0}}\ \max_{\ket s\in{\rm STAB}}\big|\bra s\,U_A\otimes U_B\ket\psi\big|^2, \label{eq:m-PhiSdef}
\end{equation}
the largest fidelity with a stabilizer state ($\ket s\in{\rm STAB}$) that parity-even local unitaries can reach; the subscript $S$ denotes this superselection constraint, which distinguishes $\Phi^{\NL}_S$ from the qubit value $\Phi^{\rm q}$. Let $a^{(\pm)}_i$ be the square roots of the Schmidt values of the two sectors, sorted in decreasing order and padded with zeros. A balanced split on $2^k$ levels is matched with the $2^{k-1}$ largest values of each sector, and a pure split with the $2^k$ largest values of one sector. Therefore \SEEAPPENDIX{(\SMref{sec:exact})}
\begin{align}
\Phi^{\NL}_S=\max\Big\{&\max_{k\ge0,\ \eta=\pm}2^{-k}\Big(\sum_{i=1}^{2^k}a^{(\eta)}_i\Big)^2,\nonumber\\
&\max_{k\ge1}2^{-k}\Big(\sum_{i=1}^{2^{k-1}}\big(a^{(+)}_i+a^{(-)}_i\big)\Big)^2\Big\}. \label{eq:m-PhiS}
\end{align}
The first line contains the pure splits, the second the balanced ones; each term is the fidelity of one flat template, the square of the sum of the amplitudes it covers divided by its number of levels [Fig.~\ref{fig:m-scheme}(c)]. A pure split needs $k+1$ modes on each side, one of them for the parity copy, and a balanced split $k$, but these limits never bind: with $n$ modes on the smaller side each sector holds at most $2^{n-1}$ nonzero values, and a term with more slots than that is smaller than the term with half as many. So $\Phi^{\NL}_S$ depends only on the graded ES, not on the numbers of modes or on unpaired modes, and below $m$ always denotes the number of BCS pairs. We measure the nonlocal magic in bits, $\Dm=-\log_2\Phi^{\NL}_S$. Compared with the qubit value $D^{\NL}_{\min}=-\log_2\Phi^{\rm q}$, superselection costs at most one bit, $\Phi^{\NL}_S\ge\frac12\Phi^{\rm q}$: a balanced template with the largest values of each sector contains the best qubit template and is at most twice as large \SEEAPPENDIX{(\SMref{sec:exact})}.

\begin{figure}[t]
\includegraphics[width=\columnwidth]{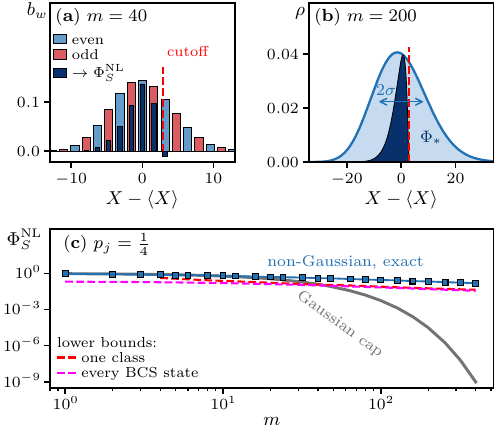}
\caption{\CHANGE{How the Bell-pair fidelity arises, against the entanglement energy $X=-\log_2\lambda$ measured from its mean. (\textbf{a}) $m=40$ identical pairs, $p=\frac14$. Light bars: weights $b_w$ of the classes of Sec.~\ref{sec:m-iid} (blue: even $w$, red: odd $w$). Dark bars: contribution of each class to the optimal fidelity $\Phi^{\NL}_S=0.40$ of Eq.~\eqref{eq:m-PhiS} ($k=32$ Bell pairs). The even template reaches past the cutoff $y+2$ of Eq.~\eqref{eq:m-kernel} (red dashed), where its levels contribute slightly negatively, a cost of parity superselection. (\textbf{b}) Haar-random Gaussian state of $N=400$ modes cut in half ($C_E=47$). Light: the density $\rho(X)$, nearly Gaussian with standard deviation $\sigma=\sqrt{C_E}/\ln2=9.9$ bits. Dark: $\rho(X)\,b(X-y^\ast)$, whose area $\Phi_\ast=0.22$ matches $4/\sqrt{2\pi(C_E+5)}$ (Sec.~\ref{sec:m-capacity}). (\textbf{c}) Fidelity for $p_j=\frac14$: exact value [Eq.~\eqref{eq:m-PhiS}], the lower bounds of Eqs.~\eqref{eq:m-oneclass} and \eqref{eq:m-conv}, and the cap $(13/16)^{m/4}$ for local Gaussian unitaries [Eq.~\eqref{eq:m-gauss}].}}\label{fig:m-fidelity}
\end{figure}

\subsection{\texorpdfstring{\CHANGE{A single pair}}{A single pair}}\label{sec:m-onepair}

For one BCS pair ($m=1$), $\sqrt q\ket{00}+\sqrt p\ket{11}$ with $p=\sin^2\theta\le\frac12$, each sector holds one Schmidt value, $a^{(+)}_1=\sqrt q$ and $a^{(-)}_1=\sqrt p$, so only the pure terms with $k=0$ and the balanced term with $k=1$ matter. Dropping the pure term $p\le q$, Eq.~\eqref{eq:m-PhiS} gives: 
\begin{align}
\Phi^{\NL}_S(m{=}1)&=\max\big\{q,\ \tfrac12(\sqrt q+\sqrt p)^2\big\}.\label{eq:m-onepair}
\end{align}
The first entry, $q$, is the fidelity with the empty state $\ket0_A\ket0_B$, whereas $\frac12+\sqrt{pq}$ is that with one fermionic Bell pair. The Bell pair is the better target for $p\ge\sin^2\frac\pi8\approx0.15$. At $p=\sin^2\frac\pi8$, the two targets tie at fidelity $\cos^2\frac\pi8\approx0.85$: the pair with the largest Gaussian-restricted magic is the hardest to convert with $\Dm\approx -\log_{2}(0.85) \approx 0.23$. For one mode on each side every parity-even unitary is Gaussian; the two answers separate only for $m \ge 3$ pairs.

The fidelity also controls the SRE: every pure state obeys $M_2\le4D_{\min}$~\cite{Haug2023Stabilizer,Sierant2026Exact}, with $D_{\min}$ the $-\log_2$ of its largest fidelity with a stabilizer state, so $M^{f,\NL}_2\le4\Dm$ \SEEAPPENDIX{(\SMref{app:exact})}. Since one $T$ gate raises $M_2$ by at most $\log_2\frac43$ \CHANGE{(Theorem~2 of Ref.~\cite{Zhu2024Amortized})}, any Clifford$+T$ circuit that prepares $\ket\psi$ exactly from a stabilizer input, up to parity-even local unitaries that are not counted, contains at least $M^{f,\NL}_2/\log_2\frac43\approx2.41\,M^{f,\NL}_2$ $T$ gates \CHANGE{(Corollary~4 there)}.

\subsection{\texorpdfstring{\CHANGE{Identical pairs: a simple picture}}{Identical pairs: a simple picture}}\label{sec:m-iid}

Consider $m$ BCS pairs with the same occupation $p<\frac12$. Each Schmidt vector is a Fock configuration $s\in\{0,1\}^m$ of the pairs, with $s_j=1$ if pair $j$ is occupied, and its Schmidt value depends only on the number $w=|s|$ of occupied pairs, $\lambda_w=p^wq^{m-w}$. \CHANGE{Its entanglement energy is $X_w=-\log_2\lambda_w=-m\log_2q+w\log_2(q/p)$, so each occupied pair raises it by the same $\log_2(q/p)$ bits.} The ES is therefore a ladder of classes: class $w$ contains $\binom mw$ equal Schmidt values, all with parity $(-1)^w$, and carries the total weight $b_w=\binom mw\lambda_w$, the binomial distribution [Fig.~\ref{fig:m-fidelity}(a)]. For large $m$ this distribution is close to a Gaussian of mean $mp$ and variance $mpq$ (de Moivre--Laplace theorem, the simplest local limit theorem), so the most likely class carries weight $\simeq1/\sqrt{2\pi mpq}$.

Local Gaussian unitaries cannot exploit these classes: their fidelity with every stabilizer state decays exponentially in $m$ (Sec.~\ref{sec:m-gaussworse}). Leaving the pairs as they are, as $m$ fermionic Bell pairs with losses that multiply, gives $(\frac12+\sqrt{pq})^m\approx0.933^m$ at $p=\frac14$ [Eq.~\eqref{eq:m-onepair}].

Non-Gaussian unitaries can do much better, because the levels of a class are all equal, exactly like those of a flat template. A template that keeps $2^{k_w}$ levels of class $w$, with $k_w=\lfloor\log_2\binom mw\rfloor$, covers more than half of the class, since $\binom mw<2^{k_w+1}$, and a local Fock-state permutation, parity-even but not Gaussian, moves these levels onto the Schmidt vectors of $k_w$ Bell pairs with their parity stored in one more mode. The fidelity $(2^{k_w}\sqrt{\lambda_w})^2/2^{k_w}=2^{k_w}\lambda_w$ is thus more than half the class weight, and the most likely class gives
\begin{equation}
\Phi^{\NL}_S\ge\max_w2^{k_w}\lambda_w>\tfrac12\max_wb_w\simeq\frac1{2\sqrt{2\pi mpq}}. \label{eq:m-oneclass}
\end{equation}
At small $m$ a single class is worse than leaving the pairs alone, but the product $(\frac12+\sqrt{pq})^m$ decays exponentially in $m$, whereas the single-class fidelity decays only as $1/\sqrt m$, like the probability of the most likely class. At $p=\frac14$ the product is the exact optimum of Eq.~\eqref{eq:m-PhiS} up to $m=5$; the optimum first exceeds it at $m=6$ (fidelity $0.72$ against $0.66$), and from $m=8$ on it exceeds the Gaussian cap, so non-Gaussian unitaries then provably beat every local Gaussian unitary [Fig.~\ref{fig:m-fidelity}(c)].

\section{Free fermions: from extensive to logarithmic}\label{sec:m-free}

We first bound the fidelity of every state by its capacity (A) and then show that for $m$ BCS pairs the capacity grows at most linearly in $m$ (B). We then compare with local Gaussian unitaries (C), translate fidelities into the SRE and $T$-gate counts (D), and give asymptotic laws for many pairs and for critical chains (E). Identical pairs with $p=\frac14$ (Sec.~\ref{sec:m-iid}) serve as a running example.

\subsection{A bound for every state}\label{sec:m-capacity}

\CHANGE{For a general state the Schmidt values are not degenerate and have no class structure, although nearly equal values can play the role of a class, so we follow their distribution statistically.} To do this, \CHANGE{regard the entanglement energy in bits, $X$ of Sec.~\ref{sec:m-background},} as a random variable that takes the value $X_i$ with probability $\lambda_i$; its mean is $S_1$ and its variance is $C_E/\ln^22$ [Eq.~\eqref{eq:m-CE}]. If we drop the two restrictions in Eq.~\eqref{eq:m-PhiS}, so that the number of levels of the template need not be a power of two and the parity sectors are ignored, the best flat template obeys the exact identity
\begin{equation}
\Phi_\ast\equiv\max_{K}\frac1K\Big(\sum_{i=1}^K\sqrt{\lambda_i}\Big)^2=\sup_{y}\,\mathbb E\,b(X-y), \label{eq:m-kernel}
\end{equation}
with \CHANGE{the Schmidt values sorted, $\lambda_1\ge\lambda_2\ge\cdots$,} and the kernel $b(z)=(2\cdot2^{z/2}-2^z)_+$; it follows by completing a square \SEEAPPENDIX{(\SMref{app:capacity})}. \CHANGE{Here $\Phi_\ast$ is the best fidelity that local unitaries reach with a maximally entangled state of any Schmidt rank $K$, which need not be a stabilizer state. In the derivation each kept level earns $2\tau\sqrt{\lambda_i}$ and pays $\tau^2$ for its slot, and at the optimum $\tau=2^{-y/2}$ is the average amplitude of the kept levels. Per unit of probability a level thus contributes $b=2r-r^2=1-(1-r)^2$, with $r=\tau/\sqrt{\lambda_i}=2^{(X_i-y)/2}$: the kernel is an exchange rate between probability and fidelity, and it depends on a level only through its energy. Hence $\Phi_\ast$ depends on the state only through the distribution of $X$, for identical pairs a histogram of $m+1$ classes rather than a list of $2^m$ levels.} The kernel is a window. A level at energy $X$ contributes $\lambda\,b(X-y)$ to the fidelity, which vanishes above the cutoff $y+2$, is largest at $X=y$, and halves every two bits further below, where the levels have larger Schmidt values but are too few to fill the template. The fidelity is thus collected from a band a few bits wide: the class picture of Sec.~\ref{sec:m-iid}, smeared over about six neighboring classes for identical pairs [Fig.~\ref{fig:m-fidelity}(a)]. Quantitatively, $\Phi_\ast$ is $4/\ln2$ times the peak of the density of $X$ smoothed over about $3.2$ bits, which raises its variance to $(C_E+5)/\ln^22$. For a nearly Gaussian $X$, as for a generic free-fermion state [Fig.~\ref{fig:m-fidelity}(b)], this gives $\Phi_\ast\simeq4/\sqrt{2\pi(C_E+5)}$, which falls as $1/\sqrt{C_E}$. \CHANGE{The window has area $\int b=4/\ln2\approx5.8$ bits, so this estimate needs $X$ to be spread over many more bits than that; a distribution narrower than the window fits into it almost entirely, and then $\Phi_\ast$ is of order one.}

A density of variance $\sigma^2$ has a peak of at least $(12\sigma^2)^{-1/2}$, the value for a uniform density, so $\Phi_\ast\ge2/\sqrt{3(C_E+5)}$ for every state. \CHANGE{Each of the two restrictions only removes templates, so $\Phi^{\NL}_S\le\Phi^{\rm q}\le\Phi_\ast$. For four identical pairs at $p=\frac14$, for example, the best template of any size holds the $11$ levels with at most two occupied pairs, with $\Phi_\ast=0.811$, whereas the best admissible one holds all $16$ levels, four fermionic Bell pairs, with $\Phi^{\NL}_S=\Phi^{\rm q}=0.758$. The loss is bounded: rounding the optimal size $K$, with $2^k\le K<2^{k+1}$, down or up to a power of two keeps at least a fraction $\max(2^k/K,K/2^{k+1})\ge1/\sqrt2$ of the fidelity, so $\Phi^{\rm q}\ge\Phi_\ast/\sqrt2$.} Restoring the power-of-two and parity restrictions costs at most a factor $\sqrt2-1$ \SEEAPPENDIX{(Proposition~\ref{prop:capbound} of the SM)}, so every pure fermionic state of definite parity obeys
\begin{equation}
\Phi^{\NL}_S\ge\frac{2(\sqrt2-1)}{\sqrt{3(C_E+5)}}\approx\frac{0.478}{\sqrt{C_E+5}}, \label{eq:m-capbound}
\end{equation}
that is, using $\Dm=-\log_2\Phi^{\NL}_S$, we have
$\Dm\le\frac12\log_2[\tfrac34(3+2\sqrt2)(C_E+5)]$. 
\emph{The nonlocal magic of every state is thus bounded, at every size, by the thermal fluctuations of its entanglement Hamiltonian}. 

\subsection{The capacity of a BCS state and a fidelity of order $1/\sqrt m$}\label{sec:m-bcs}

Start with a single pair, $\sqrt q\ket{00}+\sqrt p\ket{11}$. Its reduced density matrix on $A$ has two eigenvalues, $q$ and $p$, so the entanglement Hamiltonian $H_E=-\ln\rho_A$ has two levels, $-\ln q$ and $-\ln p$, occupied with probabilities $q$ and $p$. The capacity is the variance of this two-valued random variable. A variable that takes two values with probabilities $q$ and $p$ has variance $pq$ times the square of their difference, so the capacity of one pair is
\begin{equation}
c(p)=pq\ln^2\frac qp. \label{eq:m-cp}
\end{equation}
It vanishes in two cases. For a product pair, $p=0$, only one level is occupied and nothing fluctuates. For a Bell pair, $p=\frac12$, the two levels coincide, the ES is flat, and again nothing fluctuates. In between $c(p)$ is positive, so it has a maximum. 
\CHANGE{Setting the derivative to zero gives $\ln(q/p)=2/(1-2p)$, so at the maximum $c=4pq/(1-2p)^2$. The maximum is attained at $p\approx0.0832$ and at $1-0.0832$, and it is}
\begin{equation}
c_\ast=\max_pc(p)\approx0.439. \label{eq:m-cstar}
\end{equation}
Now take $m$ pairs. The reduced density matrix of a product of pairs is a product, $\rho_A=\bigotimes_j\rho_A^{(j)}$, so the entanglement Hamiltonian is a sum, $H_E=\sum_jH_E^{(j)}$, and in the state $\rho_A$ the terms are independent random variables. The variance of a sum of independent variables is the sum of their variances, exactly as for the number of heads in $m$ coin flips, so
\begin{equation}
C_E=\sum_{j=1}^mc(p_j)\le c_\ast m. \label{eq:m-CEbcs}
\end{equation}
For identical pairs with $p=1/4$, $C_E=0.226\,m$. \CHANGE{In bits the entanglement energy thus has standard deviation $\sqrt{C_E}/\ln2\le\sqrt{c_\ast m}/\ln2\approx0.96\sqrt m$: its mean, the entropy, grows like $m$, but its spread only like $\sqrt m$, as for the number of heads in $m$ coin flips. For identical pairs, $X$ is a constant plus $\log_2(q/p)$ times the binomial number of occupied pairs.} This linear growth is the only place where the free-fermion structure enters. A general state with $m$ modes on each side can spread its Schmidt values evenly over $m$ bits of entanglement energy, for example $\lambda_R\propto1/R$ for $R=1,\dots,2^m$. Then $C_E$ grows as $m^2$, and its fidelity can be as small as order $1/m$, which is the worst case: a dyadic-shell bound gives $\Phi^{\NL}_S\ge3(2+\sqrt2)/(8n)$, that is, $\Dm\le\log_2n-0.35$, for every state with $n\ge2$ modes on the smaller side \CHANGE{(Proposition~\ref{prop:dyadic-fidelity} of the SM)}. Independent pairs cannot spread their energy this widely: each adds at most $c_\ast$ to the variance, which halves the logarithmic rate.
Inserting $C_E\le c_\ast m = 0.439 m$ into Eq.~\eqref{eq:m-capbound}, and writing $c_\ast m+5=c_\ast(m+5/c_\ast)$, gives for every BCS state of $m$ pairs
\begin{equation}
\Phi^{\NL}_S\ge\frac{2(\sqrt2-1)}{\sqrt{3(c_\ast m+5)}}\ge\frac{0.72}{\sqrt{m+11.4}}, \label{eq:m-conv}
\end{equation}
The fidelity thus decays no faster than $1/\sqrt m$. 
\CHANGE{For partially entangled pairs it also decays no slower. The kernel vanishes above $y+2$ and halves every two bits below $y$, so $\Phi^{\NL}_S\le\Phi_\ast$ is at most a few times the largest probability that $X$ falls into an interval a few bits wide, and by the Berry--Esseen theorem this probability is of order $1/\sqrt{C_E}$. Each pair adds at least $c_\delta=\min\{c(\delta),c(\frac12-\delta)\}>0$ to $C_E$, so $\Phi^{\NL}_S\le2^{a_\delta}/\sqrt{c_\delta m}$ with a constant $a_\delta$ that depends only on $\delta$ (Proposition~\ref{prop:bcs-converse} of the SM); for $p_j\equiv\frac14$ this reads $\Phi^{\NL}_S\le10.4/\sqrt m$. Hence $\Phi^{\NL}_S=\Theta(m^{-1/2})$ and $\Dm=\frac12\log_2m+O(1)$. Pairs that are nearly Bell pairs or nearly empty must be excluded: they add almost nothing to $C_E$ and keep the fidelity close to one.}

\subsection{Local Gaussian unitaries do exponentially worse}\label{sec:m-gaussworse}

\begin{figure}[t]
\includegraphics[width=\columnwidth]{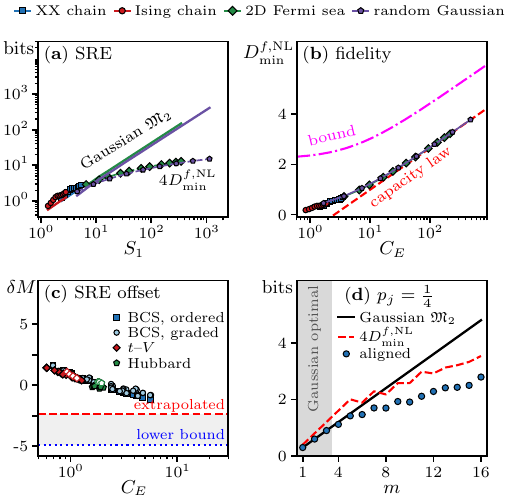}
\caption{\CHANGE{Nonlocal magic of free and interacting fermions. (\textbf{a},\,\textbf{b}) Blocks of $\ell\le8192$ sites of critical XX and Ising chains, $L\times L$ blocks ($L\le96$) of a 2D Fermi sea, and Haar-random Gaussian states of $N\le4096$ modes cut in half. (\textbf{a}) Upper bounds on the nonlocal SRE $M^{f,\NL}_2$, in bits: the Gaussian-restricted $\mathfrak M_2$ of Eq.~\eqref{eq:m-Mfree2} (solid) grows linearly with the entanglement entropy $S_1$, the certificate $4\Dm$ (dashed; Sec.~\ref{sec:m-onepair}) logarithmically. (\textbf{b}) $\Dm$ against $C_E$, with the capacity law of Eq.~\eqref{eq:m-caplaw} (red dashed, asymptotic) and the capacity bound of Eq.~\eqref{eq:m-capbound} (magenta, every state). (\textbf{c}) Offset $\delta M=M_2-\frac32\log_2C_E$ of reference states, which bound the nonlocal SRE from above (\SMref{app:small}), for random BCS states ($m\le14$) and DMRG ground states of $t$--$V$ and Hubbard chains near half filling (open markers: $V=0$, $U=0$). Graded labelings bound $M^{f,\NL}_2$, ordered ones only the qubit value $M^{\NL}_2$. Lines: asymptotic lower bound \CHANGE{of Eq.~\eqref{eq:m-sre}} ($-4.91$) and extrapolated reference-state limit i.e., $-2.39$ of the SRE offset. (\textbf{d}) Identical pairs, $p_j=\frac14$. Aligned reference states put the largest Schmidt values of each parity on the stabilizer target that attains $\Phi^{\NL}_S$; from $m=4$ on they lie below $\mathfrak M_2$. The gray shaded region shows $m\le3$, where local Gaussian unitaries are optimal (Sec.~\ref{sec:m-gaussworse}).}}\label{fig:m-laws}
\end{figure}

Now compare with local Gaussian unitaries. No local unitary changes the graded ES; what non-Gaussian unitaries add is the freedom to place the Schmidt values on the Fock states of a flat template, class by class (Sec.~\ref{sec:m-iid}). Local Gaussian unitaries lack this freedom, and the precise statement uses magic. Every state in the local-Gaussian orbit of $\ket\psi$ has SRE at least $\mathfrak M_2(p)$, the Gaussian minimum of Eq.~\eqref{eq:m-Mfree2}, by Theorem~1 of Ref.~\cite{Iannotti2026FNL}; and a pure state with SRE $M_2$ has fidelity at most $2^{-M_2/4}$ with every stabilizer state, since $M_2\le4D_{\min}$ (Sec.~\ref{sec:m-exact}). So local Gaussian unitaries reach a fidelity of at most $2^{-\mathfrak M_2(p)/4}$. This half of the separation rests entirely on that theorem, which covers every pure Gaussian state, every cut, and all local Gaussian unitaries, including those that change the particle number.

Since $\mathfrak M_2(p)$ is a sum of $m$ nonnegative terms, one per pair, it grows linearly in $m$ as long as each term is bounded away from zero. For partially entangled pairs (Sec.~\ref{sec:m-background}) the smallest term belongs to the allowed pairs closest to Bell pairs, $p_j=\frac12\pm\delta$ \CHANGE{(with $u=1-2p_j$ the term is $-\log_2(1-u^2+u^4)$, and $u^2-u^4$ is smaller at $u=2\delta$ than at $u=1-2\delta$ for $\delta\le\frac14$)}, so that
\begin{equation}
2^{-\mathfrak M_2(p)/4}\le\big[1-4\delta^2(1-4\delta^2)\big]^{m/4}, \label{eq:m-gauss}
\end{equation}
which decays exponentially in $m$. For the running example each pair contributes $\mathfrak M_2=\log_2\frac{16}{13}=0.30$ bits, so local Gaussian unitaries reach a fidelity of at most $(13/16)^{m/4}\approx0.949^m$, which is $9.6\times10^{-10}$ at $m=400$. General parity-even unitaries do far better at the same $m$: inserting $C_E=400\cdot\frac3{16}\ln^23=90.5$ into Eq.~\eqref{eq:m-capbound} guarantees a fidelity of at least $0.049$, and the exact optimum of Eq.~\eqref{eq:m-PhiS}, evaluated on the $401$ weight classes, is $0.152$.

For a few pairs the comparison is different [Fig.~\ref{fig:m-laws}(d)]. For up to three pairs no parity-even local unitary brings the SRE below $\mathfrak M_2(p)$, so local Gaussian unitaries are optimal \SEEAPPENDIX{(Proposition~\ref{thm:small-gauss} of the SM\CHANGE{, whose proof combines a parity grading of the Pauli projector with three results of Ref.~\cite{Sierant2026Spectral}})}. Superselection is essential here: among all qubit unitaries the statement already fails at three pairs. From four pairs on local Gaussian unitaries are not always optimal: for four equal pairs, a diagonal Clifford gate followed by the exchange of $\ket{1000}$ and $\ket{0111}$ on each side lowers the SRE below $\mathfrak M_2$ for every $p\notin\{0,\frac12,1\}$ \SEEAPPENDIX{(Proposition~\ref{prop:four} of the SM)}, and \CHANGE{for $p_j\equiv\frac14$ the numerical gap of Fig.~\ref{fig:m-laws}(d)} grows with $m$.

\subsection{Consequences for magic and $T$ gates}

The SRE is controlled by the fidelity through $M^{f,\NL}_2\le4\Dm$ (Sec.~\ref{sec:m-exact}). Inserting Eq.~\eqref{eq:m-capbound} with $C_E\le c_\ast m$ for the non-Gaussian side, and Eq.~\eqref{eq:m-gauss} for the Gaussian side, gives
\begin{align}
M^{f,\NL}_2&\le4\Dm\le2\log_2(1.921\,m+22),\nonumber\\
\mathfrak M_2(p)&\ge-m\log_2\big[1-4\delta^2(1-4\delta^2)\big]\nonumber\\
&\ge\tfrac{4}{\ln2}\,\delta^2(1-4\delta^2)\,m. \label{eq:m-sep}
\end{align}
The first line is Eq.~\eqref{eq:m-capbound} with $C_E\le c_\ast m$, multiplied by four, and holds for every BCS state; the second is the logarithm of Eq.~\eqref{eq:m-gauss}. Allowing non-Gaussian local unitaries thus lowers the nonlocal SRE from linear to logarithmic in $m$, and with it the lower bound $M_2/\log_2\frac43$ on the number of $T$ gates needed beyond local processing. \CHANGE{The logarithm cannot be improved: the dyadic-shell bound of Ref.~\cite{Sierant2026Spectral} gives $M^{\NL}_2\ge3D^{\NL}_{\min}-10.36$, so with the converse of Sec.~\ref{sec:m-bcs} and the one-bit cost of superselection (Sec.~\ref{sec:m-exactfid}) the nonlocal SRE of partially entangled pairs is $\Theta(\log m)$, although this lower bound becomes nontrivial only at very large $m$.} For the running example ($p=1/4$) at $m=400$, $\mathfrak M_2=119.8$ bits, so at least $119.8/\log_2(4/3) \approx 289$ $T$ gates are needed if only Gaussian processing is allowed; with general parity-even processing $M^{f,\NL}_2\le19.3$ bits \CHANGE{by Eq.~\eqref{eq:m-sep}}, so the certified lower bound drops to at most $47$ $T$ gates\CHANGE{, and the exact $\Phi^{\NL}_S=0.152$ gives $M^{f,\NL}_2\le4\Dm=10.9$ bits and at most $27$}.

\subsection{Capacity laws and critical systems}\label{sec:m-laws-par}

The capacity bound is a lower bound. For many pairs the asymptotic value of the fidelity itself can be \CHANGE{estimated}. Suppose that $C_E\to\infty$ while the third absolute moments of the pair energies stay comparable to their variances, that the energy gaps $\log_2(q_j/p_j)$ do not cluster near multiples of a common spacing \CHANGE{that stays bounded away from zero}, which identical pairs violate, and that the parity is neither frozen, as it is when almost all pairs are empty, nor locked to the energy \CHANGE{on such a lattice}, as it is on the ladder of identical pairs \CHANGE{(a spacing that shrinks, as for the critical chains below, is allowed but slows the convergence)}. Then \CHANGE{we expect} that the entanglement energy obeys a parity-resolved local limit theorem: even on the scale of one bit, which the kernel of Eq.~\eqref{eq:m-kernel} probes, $X$ is Gaussian, and every energy window is shared equally by the two parity sectors. \CHANGE{Equation~\eqref{eq:m-kernel} then gives $\Phi_\ast=4/\sqrt{2\pi C_E}\,(1+o(1))$, and if balanced stabilizer templates attain it, as we expect when the two sectors share every window, then} as $C_E\to\infty$
\begin{equation}
\Dm=\tfrac12\log_2(2\pi C_E)-2+o(1), \label{eq:m-caplaw}
\end{equation}
and the qubit value $D^{\NL}_{\min}$ obeys the same law. 
\CHANGE{Since no stabilizer fidelity exceeds $\Phi_\ast$, the right-hand side is a lower bound on both; the equality, derived for smooth qubit spectra in Ref.~\cite{Aditya2026Spreading}, we support numerically [Fig.~\ref{fig:m-laws}(b)] but do not prove.} Even the running example ($p=1/4$), which violates the second and third conditions, is close: at $m=400$ the law gives with $C_{E} = 400 (3/16) \ln^2(3) \approx 90.52$, $\Phi_\ast = 4/\sqrt{2\pi\cdot (90.52)} \approx 0.1677$, against the exact $0.152$.

For the SRE the coefficient of $\log_2C_E$ is three times larger. The fidelity probes the peak of the density $\rho$ of $X$, whose height is of order $C_E^{-1/2}$, whereas the nonlocal SRE is fixed, up to a bounded constant, by the dyadic rank shells of the ES~\cite{Sierant2026Spectral}, which probe $\int\rho^4\propto C_E^{-3/2}$: multiplying $C_E$ by $4$ costs one bit of fidelity-based magic and three bits of SRE, consistent with $M^{f,\NL}_2\le4\Dm$. \CHANGE{Quantitatively, Ref.~\cite{Sierant2026Spectral} places $M^{\NL}_2$ between $-\log_2\Qsh-\log_2C_{\rm sh}$ and $-\log_2\Qsh+4$, with $C_{\rm sh}\approx654.5$ and the shell sum $\Qsh=\lambda_1^4+\sum_{k\ge1}(2^{k-1}\lambda_{2^{k-1}+1})^4$. For Gaussian $X$, $\Qsh\simeq\int\rho^4dx/\ln^42$ and $-\log_2\Qsh=\frac32\log_2C_E+c_G+o(1)$, with $c_G=\log_2(4\sqrt2\,\pi^{3/2}\ln2)\approx4.45$. Hence, as $C_E\to\infty$,}
\begin{equation}
\begin{gathered}\CHANGE{M^{f,\NL}_2\ge M^{\NL}_2=\tfrac32\log_2C_E+O(1),}\\ \CHANGE{M^{\NL}_2\ge\tfrac32\log_2C_E-4.91+o(1),}\end{gathered} \label{eq:m-sre}
\end{equation}
\CHANGE{where $-4.91=c_G-\log_2C_{\rm sh}$. For $M^{f,\NL}_2$, Ref.~\cite{Sierant2026Spectral} gives only the lower bound, and Eq.~\eqref{eq:m-capbound} only $M^{f,\NL}_2\le4\Dm\le2\log_2C_E+O(1)$. The graded reference states [Fig.~\ref{fig:m-laws}(c)], which parity-even unitaries reach, do better: for Gaussian $X$ we estimate their offset to tend to $c_G-\log_2\mathcal K\approx-2.39$, with $\mathcal K\approx114$ the numerically extrapolated growth rate of a lattice sum. So $M^{f,\NL}_2$ very likely obeys the same law with its own offset, at accessible sizes all offsets in Fig.~\ref{fig:m-laws}(c) still lie above $-2.39$.}

Both laws use the pair structure only through the local limit theorem, so they \CHANGE{apply to} any family of fermionic states that obeys it. \CHANGE{We expect it to hold} for blocks of $\ell$ sites of critical hopping chains whose Fermi sea is a single interval of momenta, such as the XX chain. There $C_E=\frac13\ln\ell\,(1+o(1))$, so the nonlocal magic \CHANGE{should grow} only doubly logarithmically, $\Dm=\frac12\log_2(\frac{2\pi}3\ln\ell)-2+o(1)$, as found for qubit chains~\cite{Sierant2026Spectral,Torre2026NonLocal,Liu2026AntiFlatness}. This regime is reached very slowly: for $\ell\le8192$ the XX chain has $C_E\le3.53$, and at such sizes the certificate $4\Dm$ still exceeds $\mathfrak M_2$, so for critical chains the separation of Eq.~\eqref{eq:m-sep} is an asymptotic statement. It is visible at accessible sizes for states with large $C_E$: the 2D Fermi sea and Haar-random Gaussian states~\cite{Bianchi2021PageGaussian} approach Eq.~\eqref{eq:m-caplaw} to within a few hundredths of a bit at $C_E\approx100$ [Fig.~\ref{fig:m-laws}(b)], and at the largest sizes ($L=96$, $C_E=218$; $N=4096$, $C_E=479$) $4\Dm$ lies below $\mathfrak M_2$ by factors of $11.5$ and $27$ [Fig.~\ref{fig:m-laws}(a)].

\emph{Interacting fermions.} The exact fidelity of Sec.~\ref{sec:m-exact} and the capacity bound of Sec.~\ref{sec:m-capacity} hold for \emph{every} state, including interacting fermionic states. 
We evaluated both of these for DMRG ground states of $t$--$V$ chains (up to $257$ sites) and Hubbard chains (up to $64$ sites), next to half filling (methods in \SMref{sec:setting}). \CHANGE{For the SRE, the graded reference states of Eq.~\eqref{eq:ref} bound $M^{f,\NL}_2$ from above [Fig.~\ref{fig:m-laws}(c)], and the staircase bound of Ref.~\cite{Sierant2026Spectral}, Eq.~\eqref{eq:Qsh2} of the SM, bounds $M^{\NL}_2\le M^{f,\NL}_2$ from below.} For strong repulsion ($V\ge2t$, $U\ge4t$) \CHANGE{these} lower and upper bounds meet within $0.06$ bits, so in that case the nonlocal \CHANGE{SRE} of an interacting chain is essentially known exactly. At these sizes $C_E\le2.04$, and the chains look like free fermions; but so do non-integrable chains~\CHANGE{\cite{Turner2017Interaction,Matos2021Gaussianity}} and generic random spectra, so this agreement does not yet test the capacity laws, which \CHANGE{need} larger capacities.

\section{Discussion and implications}\label{sec:discussion}

\CHANGE{In this paper, we probe the question of where the \emph{magic} of a free-fermion state resides. If we do a bipartition of the state into two regions, and if each region may only be rotated by Gaussian (free-fermion) unitaries, the magic that cannot be removed (nonlocal magic) grows in proportion to the number $m$ of partially entangled pairs across the cut/bipartition. If in each region, we can apply any unitary that respects fermion parity, a Gaussian rotation followed by a reshuffling of Fock states (non-Gaussian) brings the state to Bell pairs with a fidelity that decays only as $1/\sqrt m$, and the magic left across the cut grows only like $\log m$. The state itself is a free-fermion state throughout; the local unitaries only reveal which part of its magic is local and nonlocal. Almost all the magic of a free-fermion (Gaussian) state is therefore \emph{local}: it is stored within the regions, not in the correlations between them.}
The reason is straightforward. Bell pairs have a flat entanglement spectrum, so a state can be far from Bell pairs only if its Schmidt weights are spread over many orders of magnitude, that is, only if its capacity of entanglement $C_E$, related to the variance of 
Schmidt weights is large.

\CHANGE{Two caveats apply to our results. First, we do not estimate the cost of the local unitaries, although the Fock-state permutations are themselves non-Clifford circuits. Second, magic measures the cost of a state for stabilizer (Clifford) simulation, not its classical complexity in general, so our results do not affect the efficient classical simulation of free fermions.}

\paragraph*{Fermionic quantum computation.} \CHANGE{Fermionic quantum processors~\cite{BravyiKitaev2002,GonzalezCuadra2023} manipulate fermionic modes directly, and on them parity superselection holds exactly. For operations within the parts of a register the parity-even nonlocal magic is therefore the relevant quantity. In case the hardware also conserves the particle number, the stricter number-conserving value applies, and for identical pairs its fidelity-based version stays extensive. Free-fermion gates supply the magic that makes the combination of matchgate and Clifford circuits universal~\cite{JozsaMiyake2008,BrodGalvao2011}. Our results quantify how much of that magic must be shared between two parts of a register. Across any partition, the lower bound on the number of $T$ gates that processing within the parts cannot replace grows linearly with the number $m$ of partially entangled pairs if that processing must be Gaussian, but only as $\log m$ if it may be any parity-even unitary including non-Gaussian unitaries.} 

\paragraph*{Hartree--Fock and mean-field states.} \CHANGE{Slater determinants (Hartree--Fock states) and BCS states are free-fermion states, so all our results apply to the mean-field reference states of quantum chemistry and condensed matter. These are also standard initial states of quantum algorithms~\cite{Kivlichan2018}, and their magic has recently been examined~\cite{Sarkis2025Molecules}. Our bounds need only the number $m$ of partially occupied orbital pairs across the cut, not the occupations themselves. Inserting $C_E\le c_\ast m$ into Eq.~\eqref{eq:m-capbound} gives $\Dm\le\frac12\log_2(1.921\,m+22)$ for every such state [Eq.~\eqref{eq:m-sep}].} For Hartree--Fock and BCS states, the exact $\Phi^{\NL}_S$, $C_E$, and $\mathfrak M_2$ also follow from the two-point correlations within region $A$ alone, $\langle c_i^\dagger c_j\rangle$ and $\langle c_ic_j\rangle$ with $i,j\in A$~\cite{Peschel2003}, which can be estimated efficiently from measurements~\cite{Zhao2021FermionShadows}.

\paragraph*{Classical simulation of interacting fermions.} \CHANGE{Many classical methods for interacting fermions combine a tensor network or a sampling scheme with transformations that are cheap to apply: orbital rotations, which are Gaussian unitaries, in orbital-optimized DMRG~\cite{Krumnow2016} and Gaussian matrix product states~\cite{Fishman2015}; superpositions of a few Gaussian states for quantum impurity models~\cite{BravyiGosset2017}; simulators of circuits dominated by free-fermion operations~\cite{Dias2024,ReardonSmith2024}; and Clifford circuits in Clifford-augmented matrix product states (CAMPS)~\cite{Qian2024CAMPS, Kemper:2026iaf} and stabilizer-rank methods~\cite{Bravyi2019LowRank}. Our results explains what such transformations can achieve across a single cut. Local Gaussian rotations 
cannot bring the SRE below the Gaussian-restricted nonlocal SRE $\mathfrak M_2$, whereas local Fock-state permutations reduce it to $O(\log(C_E+5))$. This does not make tensor networks cheaper: no local unitary changes the Schmidt spectrum, so the bond dimension needed at the cut stays the same. It can help only methods whose cost grows with magic, such as stabilizer-rank methods. Moreover, the permutations are not Clifford circuits, so pushing Pauli observables through them, as Clifford-augmented methods do, is in general expensive, and efficient circuits for them are not known in general. Whether permutation-augmented methods can be made efficient is open. For interacting fermions, Eq.~\eqref{eq:m-capbound} still bounds the nonlocal magic of every state by its $C_E$, from critical chains~\cite{Fang2026Microscopy} to Sachdev--Ye--Kitaev models~\cite{Bera2025SYK,Jasser2025SYK}.}

\paragraph*{Other connections.} \CHANGE{The capacity $C_E$ follows from the entanglement energies, which entanglement-Hamiltonian tomography measures in quantum simulators~\cite{Dalmonte2018,Kokail2021}. Equation~\eqref{eq:m-capbound} is thus, in principle, an experimentally accessible bound on the nonlocal magic of any pure fermionic state, including the fermionic matter of lattice gauge theories, whose magic and non-Gaussianity are now being studied~\cite{Santra2025LGT,Jha:2026ror,Jha:2026eni}. In holography, nonlocal magic has been tied to gravitational backreaction~\cite{Cao2025Gravitational}, and $C_E$ has been used as its proxy~\cite{Grieninger2026Schwinger}. 
The qubit form of Eq.~\eqref{eq:m-capbound} makes this link quantitative: the fidelity-based nonlocal magic is at most $\frac12\log_2C_E$ plus a constant, so a capacity of order $1/G_N$, with $G_N$ Newton's constant, allows only $O(\log(1/G_N))$ bits of nonlocal magic.}

\paragraph*{Outlook.} \CHANGE{The capacity of entanglement (a single number) obtained from the variance of entanglement energies bounds the nonlocal magic of every state. Whatever magic a state carries, the part that must be shared between two regions grows at most logarithmically with the fluctuation $C_E$ of its entanglement Hamiltonian. For free fermions this fluctuation grows at most linearly in the number of entangled pairs. Three questions remain open. (1)~\emph{Interacting matter.} The bound of Eq.~\eqref{eq:m-capbound} holds for every state, but whether the capacity laws, including the slope $\frac32$ of the SRE law, persist for interacting systems at large $C_E$ is not known. (2)~\emph{The cost of the local operations.} The Fock-state permutations that reduce the nonlocal magic are themselves neither Gaussian nor Clifford. A complete accounting must therefore weigh the magic that these permutations shift from the cut into the regions against their own cost: their non-Gaussianity, for which computable measures and experimentally accessible bounds were recently developed~\cite{Lakkaraju2026NonGaussianity}, and the $T$ gates they need within each region. Efficient parity-respecting circuits for them are still to be found, in particular for pairs with unequal occupations. (3)~\emph{Particle-number conservation.} 
Without a phase reference~\cite{Bartlett2007RefFrames} (a shared reservoir, such as a superconductor, with which both region $A$ and $B$ can coherently exchange particles), the local unitaries must also conserve the particle number in each region, not only its parity. For identical pairs the fidelity-based nonlocal magic then stays extensive, at least for stabilizer targets of definite particle number (Proposition~\ref{prop:u1} of the SM). However, the behavior of SRE, and both measures (fidelity-based and SRE) for unequal pairs, is still open to our knowledge. Two constants also remain to be pinned down: the best prefactor in the capacity bound, Eq.~\eqref{eq:m-capbound}, and the offset of the SRE law, Eq.~\eqref{eq:m-sre}, which we bound from below by $-4.91$ and, through reference states, estimate to be at most about $-2.4$. We leave these for future work.}

\paragraph*{Note added.} \CHANGE{While this work was being completed, a paper was posted on arXiv~\cite{Magni2026Higher} studying the Gaussian-restricted nonlocal SRE $\mathfrak M_2$ of Eq.~\eqref{eq:m-Mfree2} in higher-dimensional free-fermion systems. Both our work and theirs highlight the capacity of entanglement, but under different sets of local operations. Unlike Ref.~\cite{Magni2026Higher}, we allow non-Gaussian, parity-respecting local unitaries and show that the nonlocal magic of every pure fermionic state grows at most logarithmically with $C_E$ [Eq.~\eqref{eq:m-capbound}]. Furthermore, Ref.~\cite{Magni2026Higher} does not consider parity superselection, stabilizer fidelity, or the optimality of local Gaussian unitaries for up to three pairs, all of which are key components of the present work. In addition, another paper~\cite{Aditya2026Spreading} posted earlier in the same month, connected nonlocal magic to the capacity of entanglement for qubits and used this link to study nonlocal magic in chaotic and integrable dynamics, with consequences for entanglement embezzlement. Its saturation at a value that grows logarithmically with the subsystem size in integrable dynamics is consistent with the logarithmic laws found here. Our capacity bound, Eq.~\eqref{eq:m-capbound}, is tighter than theirs to our understanding, but Ref.~\cite{Aditya2026Spreading} considers neither parity superselection nor the optimality of local Gaussian unitaries at small sizes.} 

\section*{Acknowledgments}

We thank Norman Hogan and Heba A. Labib for comments on the draft. R.G.J., D.L., A.F.K., and B.N.B. were supported by the U.S. Department of Energy, Advanced
Scientific Computing Research, under contract number DE-SC0025384. The authors acknowledge assistance of
Anthropic's Claude Fable 5.0 and Opus 5.5 in preparing this manuscript especially in the kernel approach of Sec.~\ref{sec:m-free} and OpenAI's Astra for proving the Gaussian optimality for $m=3$ pairs. 
The authors have checked and assume complete responsibility.

\section*{Data availability}
The code and data that support the findings of this submission can be obtained from the corresponding author after publication. We include \textsc{Mathematica} notebook with this submission to verify some results in this paper. 

\bibliography{refs,kemperlab}

@article{kokcu2022algebraic,
  title={Algebraic compression of quantum circuits for Hamiltonian evolution},
  author={K{\"o}kc{\"u}, Efekan and Camps, Daan and Bassman Oftelie, Lindsay and Freericks, James K and de Jong, Wibe A and Van Beeumen, Roel and Kemper, Alexander F},
  journal={Phys. Rev. A},
  volume={105},
  number={3},
  pages={032420},
  year={2022},
  publisher={APS},
  url={https://doi.org/10.1103/PhysRevA.105.032420}
}

@article{camps2022algebraic,
  title={An algebraic quantum circuit compression algorithm for {H}amiltonian simulation},
  author={Camps, Daan and K\"okc\"u, Efekan and Bassman Oftelie, Lindsay and De Jong, Wibe A and Kemper, Alexander F and Van Beeumen, Roel},
  journal={SIAM Journal on Matrix Analysis and Applications},
  volume={43},
  number={3},
  pages={1084--1108},
  year={2022},
  publisher={SIAM},
  url={https://doi.org/10.1137/21M1439298}
}

@misc{Jha:2026ror,
    author = "Jha, Raghav G. and Toga, Goksu C. and Taher, Jaber I. and Bakalov, Bojko N. and Kemper, Alexander F.",
    title = "{Magic and entanglement in 1+1-dimensional SU(2) lattice gauge theory}",
    eprint = "2606.09971",
    archivePrefix = "arXiv",
    primaryClass = "quant-ph",
    month = "6",
    year = "2026"
}

@misc{Jha:2026eni,
    author = "Jha, Raghav G. and Taher, Jaber I. and Asaduzzaman, Muhammad and Toga, Goksu C. and Bakalov, Bojko N. and Kemper, Alexander F.",
    title = "{Non-stabilizerness and entanglement in $(2+1)$-dimensional SU(2) lattice gauge theory using tensor networks}",
    eprint = "2609.28634",
    archivePrefix = "arXiv",
    primaryClass = "quant-ph",
    month = "9",
    year = "2026"
}

@article{Bravyi2019LowRank,
  author        = {Bravyi, Sergey and Browne, Dan and Calpin, Padraic and Campbell, Earl and Gosset, David and Howard, Mark},
  title         = {Simulation of quantum circuits by low-rank stabilizer decompositions},
  journal       = {Quantum},
  volume        = {3},
  pages         = {181},
  year          = {2019},
  doi           = {10.22331/q-2019-09-02-181},
  eprint        = {1808.00128},
  archivePrefix = {arXiv}
}

@article{WWW1952,
  author        = {Wick, G. C. and Wightman, A. S. and Wigner, E. P.},
  title         = {The Intrinsic Parity of Elementary Particles},
  journal       = {Phys. Rev.},
  volume        = {88},
  pages         = {101--105},
  year          = {1952},
  doi           = {10.1103/PhysRev.88.101}
}

@article{Banuls2007Fermionic,
  author        = {Ba{\~n}uls, Mari-Carmen and Cirac, J. Ignacio and Wolf, Michael M.},
  title         = {Entanglement in fermionic systems},
  journal       = {Phys. Rev. A},
  volume        = {76},
  pages         = {022311},
  year          = {2007},
  doi           = {10.1103/PhysRevA.76.022311},
  eprint        = {0705.1103},
  archivePrefix = {arXiv}
}

@misc{Santra2025LGT,
  author        = {Santra, Gopal Chandra and Mildenberger, Julius and Ballini, Edoardo and Bottarelli, Alberto and Wauters, Matteo M. and Hauke, Philipp},
  title         = {Quantum Resources in Non-{A}belian Lattice Gauge Theories: Nonstabilizerness, Multipartite Entanglement, and Fermionic Non-{G}aussianity},
  year          = {2025},
  eprint        = {2510.07385},
  archivePrefix = {arXiv}
}

@article{Hebenstreit2019,
  author        = {Hebenstreit, Martin and Jozsa, Richard and Kraus, Barbara and Strelchuk, Sergii and Yoganathan, Mithuna},
  title         = {All pure fermionic non-{G}aussian states are magic states for matchgate computations},
  journal       = {Phys. Rev. Lett.},
  volume        = {123},
  pages         = {080503},
  year          = {2019},
  doi           = {10.1103/PhysRevLett.123.080503},
  eprint        = {1905.08584},
  archivePrefix = {arXiv}
}

@article{Dias2024,
  author        = {Dias, Beatriz and Koenig, Robert},
  title         = {Classical simulation of non-{G}aussian fermionic circuits},
  journal       = {Quantum},
  volume        = {8},
  pages         = {1350},
  year          = {2024},
  doi           = {10.22331/q-2024-05-21-1350},
  eprint        = {2307.12912},
  archivePrefix = {arXiv}
}

@article{ReardonSmith2024,
  author        = {Reardon-Smith, Oliver and Oszmaniec, Micha{\l} and Korzekwa, Kamil},
  title         = {Improved simulation of quantum circuits dominated by free fermionic operations},
  journal       = {Quantum},
  volume        = {8},
  pages         = {1549},
  year          = {2024},
  doi           = {10.22331/q-2024-12-04-1549},
  eprint        = {2307.12702},
  archivePrefix = {arXiv}
}

@article{Collura2026Gaussian,
  author        = {Collura, Mario and De Nardis, Jacopo and Alba, Vincenzo and Lami, Guglielmo},
  title         = {The non-stabilizerness of fermionic {G}aussian states},
  journal       = {Quantum},
  volume        = {10},
  pages         = {2036},
  year          = {2026},
  doi           = {10.22331/q-2026-03-23-2036},
  eprint        = {2412.05367},
  archivePrefix = {arXiv}
}

@misc{Turkeshi2026RandomGaussian,
  author        = {Turkeshi, Xhek and Sierant, Piotr and Tarabunga, Poetri Sonya},
  title         = {Pauli spectrum and nonstabilizerness of random fermionic {G}aussian states},
  year          = {2026},
  eprint        = {2609.15739},
  archivePrefix = {arXiv}
}

@misc{Iannotti2026FNL,
  author        = {Iannotti, Daniele and Magni, Beatrice and Cioli, Riccardo and Hamma, Alioscia and Turkeshi, Xhek},
  title         = {Non-Local Magic Resources for Fermionic {G}aussian States},
  year          = {2026},
  eprint        = {2604.27049v3},
  archivePrefix = {arXiv}
}

@misc{Collura2026NLFF,
  author        = {Collura, Mario and B{\'e}ri, Benjamin and Tirrito, Emanuele},
  title         = {Nonlocal nonstabilizerness in free fermion models},
  year          = {2026},
  eprint        = {2604.27055},
  archivePrefix = {arXiv}
}

@article{Leone2022SRE,
  author        = {Leone, Lorenzo and Oliviero, Salvatore F. E. and Hamma, Alioscia},
  title         = {Stabilizer {R}{\'e}nyi Entropy},
  journal       = {Phys. Rev. Lett.},
  volume        = {128},
  pages         = {050402},
  year          = {2022},
  doi           = {10.1103/PhysRevLett.128.050402}
}

@article{Cao2025Gravitational,
  author        = {Cao, ChunJun and Cheng, Gong and Hamma, Alioscia and Leone, Lorenzo and Munizzi, William and Oliviero, Salvatore F. E.},
  title         = {Gravitational back-reaction is magical},
  journal       = {PRX Quantum},
  volume        = {6},
  pages         = {040375},
  year          = {2025},
  doi           = {10.1103/z3vr-w5c5},
  eprint        = {2403.07056},
  archivePrefix = {arXiv}
}

@article{Tirrito2024Flatness,
  author        = {Tirrito, Emanuele and Tarabunga, Poetri Sonya and Lami, Guglielmo and Chanda, Titas and Leone, Lorenzo and Oliviero, Salvatore F. E. and Dalmonte, Marcello and Collura, Mario and Hamma, Alioscia},
  title         = {Quantifying non-stabilizerness through entanglement spectrum flatness},
  journal       = {Phys. Rev. A},
  volume        = {109},
  pages         = {L040401},
  year          = {2024},
  doi           = {10.1103/PhysRevA.109.L040401},
  eprint        = {2304.01175},
  archivePrefix = {arXiv}
}

@misc{Sierant2026Exact,
  author        = {Sierant, Piotr},
  title         = {Exact quantification of nonlocal magic},
  year          = {2026},
  eprint        = {2608.28563},
  archivePrefix = {arXiv}
}

@misc{Viscardi2026Purification,
  author        = {Viscardi, Michele and Leone, Lorenzo and Hamma, Alioscia},
  title         = {Non-local Magic: closed-form solution and equivalence with magic of purification},
  year          = {2026},
  eprint        = {2609.04119},
  archivePrefix = {arXiv}
}

@misc{Sierant2026Spectral,
  author        = {Sierant, Piotr},
  title         = {Spectral geometry of nonlocal stabilizer entropy},
  year          = {2026},
  eprint        = {2609.13106},
  archivePrefix = {arXiv}
}

@misc{Aditya2026Spreading,
  author        = {Aditya, Sreemayee and Sierant, Piotr and Turkeshi, Xhek},
  title         = {Nonlocal Magic Spreading in Many-body Quantum Dynamics: From Chaotic Evolution to Quasi-particle Picture in Integrable Models},
  year          = {2026},
  eprint        = {2609.20951},
  archivePrefix = {arXiv}
}

@misc{Torre2026NonLocal,
  author        = {Torre, Gianpaolo and Franchini, Fabio and Giampaolo, Salvatore Marco},
  title         = {Non-Local Magic from the Entanglement Spectrum},
  year          = {2026},
  eprint        = {2607.07808},
  archivePrefix = {arXiv}
}

@misc{Liu2026AntiFlatness,
  author        = {Liu, Lei-Yi-Nan and Cui, Jian},
  title         = {Entanglement, anti-flatness, and nonlocal nonstabilizerness: a unified perspective from entanglement spectrum},
  year          = {2026},
  eprint        = {2609.01993},
  archivePrefix = {arXiv}
}

@article{Bera2025SYK,
  author        = {Bera, Surajit and Schir{\`o}, Marco},
  title         = {Non-Stabilizerness of {S}achdev-{Y}e-{K}itaev Model},
  journal       = {SciPost Phys.},
  volume        = {19},
  pages         = {159},
  year          = {2025},
  doi           = {10.21468/SciPostPhys.19.6.159},
  eprint        = {2502.01582},
  archivePrefix = {arXiv}
}

@article{Jasser2025SYK,
  author        = {Jasser, Barbara and Odavi{\'c}, Jovan and Hamma, Alioscia},
  title         = {Stabilizer Entropy and entanglement complexity in the {S}achdev-{Y}e-{K}itaev model},
  journal       = {Phys. Rev. B},
  volume        = {112},
  pages         = {174204},
  year          = {2025},
  doi           = {10.1103/rz86-47h3},
  eprint        = {2502.03093},
  archivePrefix = {arXiv}
}

@misc{Sarkis2025Molecules,
  author        = {Sarkis, Matthieu and Tkatchenko, Alexandre},
  title         = {Are Molecules Magical? {N}on-Stabilizerness in Molecular Bonding},
  year          = {2025},
  eprint        = {2504.06673},
  archivePrefix = {arXiv}
}

@misc{Fang2026Microscopy,
  author        = {Fang, Jun Qi and Wang, Fo-Hong and Xu, Xiao Yan},
  title         = {Stabilizer-{R}{\'e}nyi Microscopy of Critical Correlations in Interacting Fermions},
  year          = {2026},
  eprint        = {2601.13314},
  archivePrefix = {arXiv}
}

@article{BravyiKitaev2002,
  author        = {Bravyi, Sergey B. and Kitaev, Alexei Yu.},
  title         = {Fermionic quantum computation},
  journal       = {Ann. Phys.},
  volume        = {298},
  pages         = {210},
  year          = {2002},
  doi           = {10.1006/aphy.2002.6254}
}

@article{Botero2004BCS,
  author        = {Botero, Alonso and Reznik, Benni},
  title         = {{BCS}-like modewise entanglement of fermion {G}aussian states},
  journal       = {Phys. Lett. A},
  volume        = {331},
  pages         = {39},
  year          = {2004},
  eprint        = {quant-ph/0404176},
  archivePrefix = {arXiv}
}

@article{Yao2010Capacity,
  author        = {Yao, Hong and Qi, Xiao-Liang},
  title         = {Entanglement entropy and entanglement spectrum of the {K}itaev model},
  journal       = {Phys. Rev. Lett.},
  volume        = {105},
  pages         = {080501},
  year          = {2010},
  doi           = {10.1103/PhysRevLett.105.080501},
  eprint        = {1001.1165},
  archivePrefix = {arXiv}
}

@article{Edmonds1965,
  author        = {Edmonds, Jack},
  title         = {Maximum matching and a polyhedron with 0,1-vertices},
  journal       = {J. Res. Natl. Bur. Stand.},
  volume        = {69B},
  pages         = {125},
  year          = {1965}
}

@misc{Gottesman:1998hu,
  author        = {Gottesman, Daniel},
  title         = {The {H}eisenberg Representation of Quantum Computers},
  year          = {1998},
  eprint        = {quant-ph/9807006},
  archivePrefix = {arXiv}
}

@article{Grieninger2026Schwinger,
  author        = {Grieninger, Sebastian},
  title         = {Nonlocal nonstabilizerness from holographic {S}chwinger pair production},
  journal       = {Phys. Rev. D},
  volume        = {114},
  pages         = {L031502},
  year          = {2026},
  doi           = {10.1103/p43j-kz83},
  eprint        = {2605.04210},
  archivePrefix = {arXiv}
}

@misc{Zhu2024Amortized,
  author        = {Zhu, Chengkai and Chen, Yu-Ao and Shen, Zanqiu and Liu, Zhiping and Yu, Zhan and Wang, Xin},
  title         = {Amortized Stabilizer {R}\'enyi Entropy of Quantum Dynamics},
  year          = {2024},
  eprint        = {2409.06659},
  archivePrefix = {arXiv}
}

@misc{Magni2026Higher,
  author        = {Magni, Beatrice and Iannotti, Daniele and Travaglino, Riccardo},
  title         = {Entanglement spectrum and magic in higher-dimensional free fermionic systems},
  year          = {2026},
  eprint        = {2609.36008},
  archivePrefix = {arXiv}
}

@article{deBoer2019Capacity,
  author  = {de Boer, Jan and J{\"a}rvel{\"a}, Jarkko and Keski-Vakkuri, Esko},
  title   = {Aspects of capacity of entanglement},
  journal = {Phys. Rev. D},
  volume  = {99},
  pages   = {066012},
  year    = {2019},
  doi     = {10.1103/PhysRevD.99.066012},
  eprint  = {1807.07357},
  archivePrefix = {arXiv}
}

@article{Bao2021NonlocalMagic,
  author  = {Bao, Ning and Cao, ChunJun and Su, Vincent Paul},
  title   = {Magic state distillation from entangled states},
  journal = {Phys. Rev. A},
  volume  = {105},
  pages   = {022602},
  year    = {2022},
  doi     = {10.1103/PhysRevA.105.022602},
  eprint  = {2106.12591},
  archivePrefix = {arXiv}
}

@article{Bianchi2021PageGaussian,
  author  = {Bianchi, Eugenio and Hackl, Lucas and Kieburg, Mario},
  title   = {Page curve for fermionic {G}aussian states},
  journal = {Phys. Rev. B},
  volume  = {103},
  pages   = {L241118},
  year    = {2021},
  doi     = {10.1103/PhysRevB.103.L241118},
  eprint  = {2103.05416},
  archivePrefix = {arXiv}
}

@article{LiHaldane2008,
  author        = {Li, Hui and Haldane, F. D. M.},
  title         = {Entanglement Spectrum as a Generalization of Entanglement Entropy: Identification of Topological Order in Non-{A}belian Fractional Quantum {H}all Effect States},
  journal       = {Phys. Rev. Lett.},
  volume        = {101},
  pages         = {010504},
  year          = {2008},
  doi           = {10.1103/PhysRevLett.101.010504}
}

@article{Peschel2003,
  author        = {Peschel, Ingo},
  title         = {Calculation of reduced density matrices from correlation functions},
  journal       = {J. Phys. A: Math. Gen.},
  volume        = {36},
  pages         = {L205},
  year          = {2003},
  doi           = {10.1088/0305-4470/36/14/101}
}

@article{Dalmonte2018,
  author        = {Dalmonte, M. and Vermersch, B. and Zoller, P.},
  title         = {Quantum simulation and spectroscopy of entanglement {H}amiltonians},
  journal       = {Nat. Phys.},
  volume        = {14},
  pages         = {827},
  year          = {2018},
  doi           = {10.1038/s41567-018-0151-7}
}

@article{Kokail2021,
  author        = {Kokail, Christian and van Bijnen, Rick and Elben, Andreas and Vermersch, Beno{\^\i}t and Zoller, Peter},
  title         = {Entanglement {H}amiltonian tomography in quantum simulation},
  journal       = {Nat. Phys.},
  volume        = {17},
  pages         = {936},
  year          = {2021},
  doi           = {10.1038/s41567-021-01260-w}
}

@article{BravyiKitaev2005,
  author        = {Bravyi, Sergey and Kitaev, Alexei},
  title         = {Universal quantum computation with ideal {C}lifford gates and noisy ancillas},
  journal       = {Phys. Rev. A},
  volume        = {71},
  pages         = {022316},
  year          = {2005},
  doi           = {10.1103/PhysRevA.71.022316}
}

@article{JozsaMiyake2008,
  author        = {Jozsa, Richard and Miyake, Akimasa},
  title         = {Matchgates and classical simulation of quantum circuits},
  journal       = {Proc. R. Soc. A},
  volume        = {464},
  pages         = {3089},
  year          = {2008},
  doi           = {10.1098/rspa.2008.0189},
  eprint        = {0804.4050},
  archivePrefix = {arXiv}
}

@article{BrodGalvao2011,
  author        = {Brod, Daniel J. and Galv{\~a}o, Ernesto F.},
  title         = {Extending matchgates into universal quantum computation},
  journal       = {Phys. Rev. A},
  volume        = {84},
  pages         = {022310},
  year          = {2011},
  doi           = {10.1103/PhysRevA.84.022310},
  eprint        = {1106.1863},
  archivePrefix = {arXiv}
}

@misc{Projansky2024,
  author        = {Projansky, Andrew M. and Necaise, Jason and Whitfield, James D.},
  title         = {Gaussianity and Simulability of {C}liffords and Matchgates},
  year          = {2024},
  eprint        = {2410.10068},
  archivePrefix = {arXiv}
}

@article{Zhao2021FermionShadows,
  author        = {Zhao, Andrew and Rubin, Nicholas C. and Miyake, Akimasa},
  title         = {Fermionic partial tomography via classical shadows},
  journal       = {Phys. Rev. Lett.},
  volume        = {127},
  pages         = {110504},
  year          = {2021},
  doi           = {10.1103/PhysRevLett.127.110504},
  eprint        = {2010.16094},
  archivePrefix = {arXiv}
}

@article{Jiang2018Gaussian,
  author        = {Jiang, Zhang and Sung, Kevin J. and Kechedzhi, Kostyantyn and Smelyanskiy, Vadim N. and Boixo, Sergio},
  title         = {Quantum algorithms to simulate many-body physics of correlated fermions},
  journal       = {Phys. Rev. Appl.},
  volume        = {9},
  pages         = {044036},
  year          = {2018},
  doi           = {10.1103/PhysRevApplied.9.044036},
  eprint        = {1711.05395},
  archivePrefix = {arXiv}
}

@article{RossSelinger2016,
  author        = {Ross, Neil J. and Selinger, Peter},
  title         = {Optimal ancilla-free {C}lifford+{T} approximation of $z$-rotations},
  journal       = {Quantum Inf. Comput.},
  volume        = {16},
  pages         = {901},
  year          = {2016},
  eprint        = {1403.2975},
  archivePrefix = {arXiv}
}

@article{Bartlett2007RefFrames,
  author        = {Bartlett, Stephen D. and Rudolph, Terry and Spekkens, Robert W.},
  title         = {Reference frames, superselection rules, and quantum information},
  journal       = {Rev. Mod. Phys.},
  volume        = {79},
  pages         = {555},
  year          = {2007},
  doi           = {10.1103/RevModPhys.79.555}
}

@article{EastinKnill2009,
  author  = {Eastin, Bryan and Knill, Emanuel},
  title   = {Restrictions on Transversal Encoded Quantum Gate Sets},
  journal = {Phys. Rev. Lett.},
  volume  = {102},
  pages   = {110502},
  year    = {2009},
  doi     = {10.1103/PhysRevLett.102.110502},
  eprint  = {0811.4262},
  archivePrefix = {arXiv}
}

@article{Litinski2019,
  author  = {Litinski, Daniel},
  title   = {Magic State Distillation: Not as Costly as You Think},
  journal = {Quantum},
  volume  = {3},
  pages   = {205},
  year    = {2019},
  doi     = {10.22331/q-2019-12-02-205},
  eprint  = {1905.06903},
  archivePrefix = {arXiv}
}

@article{Kivlichan2018,
  author        = {Kivlichan, Ian D. and McClean, Jarrod and Wiebe, Nathan and Gidney, Craig and Aspuru-Guzik, Al{\'a}n and Chan, Garnet Kin-Lic and Babbush, Ryan},
  title         = {Quantum Simulation of Electronic Structure with Linear Depth and Connectivity},
  journal       = {Phys. Rev. Lett.},
  volume        = {120},
  pages         = {110501},
  year          = {2018},
  doi           = {10.1103/PhysRevLett.120.110501}
}

@article{Qian2024CAMPS,
  author        = {Qian, Xiangjian and Huang, Jiale and Qin, Mingpu},
  title         = {Augmenting Density Matrix Renormalization Group with {C}lifford Circuits},
  journal       = {Phys. Rev. Lett.},
  volume        = {133},
  pages         = {190402},
  year          = {2024},
  doi           = {10.1103/PhysRevLett.133.190402}
}

@article{Krumnow2016,
  author        = {Krumnow, C. and Veis, L. and Legeza, {\"O}. and Eisert, J.},
  title         = {Fermionic Orbital Optimization in Tensor Network States},
  journal       = {Phys. Rev. Lett.},
  volume        = {117},
  pages         = {210402},
  year          = {2016},
  doi           = {10.1103/PhysRevLett.117.210402}
}

@article{Fishman2015,
  author        = {Fishman, Matthew T. and White, Steven R.},
  title         = {Compression of correlation matrices and an efficient method for forming matrix product states of fermionic {G}aussian states},
  journal       = {Phys. Rev. B},
  volume        = {92},
  pages         = {075132},
  year          = {2015},
  doi           = {10.1103/PhysRevB.92.075132}
}

@article{BravyiGosset2017,
  author        = {Bravyi, Sergey and Gosset, David},
  title         = {Complexity of Quantum Impurity Problems},
  journal       = {Commun. Math. Phys.},
  volume        = {356},
  pages         = {451--500},
  year          = {2017},
  doi           = {10.1007/s00220-017-2976-9}
}

@article{GonzalezCuadra2023,
  author        = {Gonz{\'a}lez-Cuadra, D. and Bluvstein, D. and Kalinowski, M. and Kaubruegger, R. and Maskara, N. and Naldesi, P. and Zache, T. V. and Kaufman, A. M. and Lukin, M. D. and Pichler, H. and others},
  title         = {Fermionic quantum processing with programmable neutral atom arrays},
  journal       = {Proc. Natl. Acad. Sci. USA},
  volume        = {120},
  pages         = {e2304294120},
  year          = {2023},
  doi           = {10.1073/pnas.2304294120}
}

@article{Turner2017Interaction,
  author        = {Turner, Christopher J. and Meichanetzidis, Konstantinos and Papi{\'c}, Zlatko and Pachos, Jiannis K.},
  title         = {Optimal free descriptions of many-body theories},
  journal       = {Nat. Commun.},
  volume        = {8},
  pages         = {14926},
  year          = {2017},
  doi           = {10.1038/ncomms14926}
}

@article{Matos2021Gaussianity,
  author        = {Matos, Gabriel and Hallam, Andrew and Deger, Aydin and Papi{\'c}, Zlatko and Pachos, Jiannis K.},
  title         = {Emergence of {G}aussianity in the thermodynamic limit of interacting fermions},
  journal       = {Phys. Rev. B},
  volume        = {104},
  pages         = {L180408},
  year          = {2021},
  doi           = {10.1103/PhysRevB.104.L180408}
}

@misc{Lakkaraju2026NonGaussianity,
  author        = {Lakkaraju, Leela Ganesh Chandra and Varikuti, Naga Dileep and Hauke, Philipp},
  title         = {Fermionic non-{G}aussianity via computable measures and experimentally accessible bounds},
  year          = {2026},
  eprint        = {2610.17219},
  archivePrefix = {arXiv}
}

@misc{Kemper:2026iaf,
    author = "Kemper, Alexander F. and Jha, Raghav G. and Bachhar, Arnab and Toga, Goksu C. and Perez, Mariano Guerrero and Mayhall, Nicholas J.",
    title = "{Efficient computation of real-time correlators using Pauli Propagation}",
    eprint = "2607.24924",
    archivePrefix = "arXiv",
    primaryClass = "quant-ph",
    month = "7",
    year = "2026"
}

@article{Haug2023Stabilizer,
  author        = {Haug, Tobias and Piroli, Lorenzo},
  title         = {Stabilizer entropies and nonstabilizerness monotones},
  journal       = {Quantum},
  volume        = {7},
  pages         = {1092},
  year          = {2023},
  doi           = {10.22331/q-2023-08-28-1092}
}

@article{hogan2026efficient,
  author        = {Hogan, Norman and K{\"o}kc{\"u}, Efekan and Steckmann, Thomas and Doak, Liam P. and Mejuto-Zaera, Carlos and Camps, Daan and Van Beeumen, Roel and de Jong, Wibe A. and Kemper, A. F.},
  title         = {Efficient quantum implementation of dynamical mean field theory for correlated materials},
  journal       = {npj Comput. Mater.},
  year          = {2026},
  doi           = {10.1038/s41524-026-02289-2}
}

\clearpage
\onecolumngrid
\begin{center}
{\large\bfseries Supplemental Material for ``Nonlocal magic of fermions is bounded by entanglement fluctuations and is logarithmic for free fermions''}\\[4pt]
Raghav G.~Jha, Dengyao Luo, Alexander F.~Kemper, and Bojko N.~Bakalov
\end{center}
\vspace{6pt}
\twocolumngrid
\setcounter{section}{0}\setcounter{equation}{0}\setcounter{figure}{0}\setcounter{table}{0}
\setcounter{theorem}{0}\setcounter{lemma}{0}\setcounter{proposition}{0}\setcounter{corollary}{0}\setcounter{conjecture}{0}
\renewcommand{\thesection}{S\arabic{section}}\renewcommand{\theHsection}{S\arabic{section}}
\renewcommand{\theequation}{S\arabic{equation}}\renewcommand{\theHequation}{S\arabic{equation}}
\renewcommand{\thefigure}{S\arabic{figure}}\renewcommand{\theHfigure}{S\arabic{figure}}
\renewcommand{\thetable}{S\arabic{table}}\renewcommand{\theHtable}{S\arabic{table}}
\renewcommand{\thetheorem}{S\arabic{theorem}}\renewcommand{\theHtheorem}{S\arabic{theorem}}
\renewcommand{\thelemma}{S\arabic{lemma}}\renewcommand{\theHlemma}{S\arabic{lemma}}
\renewcommand{\theproposition}{S\arabic{proposition}}\renewcommand{\theHproposition}{S\arabic{proposition}}
\renewcommand{\thecorollary}{S\arabic{corollary}}\renewcommand{\theHcorollary}{S\arabic{corollary}}
\renewcommand{\theconjecture}{S\arabic{conjecture}}\renewcommand{\theHconjecture}{S\arabic{conjecture}}

This Supplemental Material (SM) contains the
proofs and numerical methods.
Section~\ref{sec:setting} fixes the notation and
summarizes the Gaussian-restricted result.
Section~\ref{sec:exact} derives the exact
parity-resolved stabilizer fidelity, the capacity
bound, and the finite-size estimates.
Section~\ref{sec:small} proves Gaussian optimality
for up to three BCS pairs and gives four-pair
counterexamples.
Section~\ref{sec:remarks} treats local
particle-number conservation.

\section{Setting, notation, and prior work}\label{sec:setting}

\paragraph*{Setting and conventions.}
We consider pure fermionic states of definite total parity on
$N=N_A+N_B$ modes, with the Jordan--Wigner (JW) ordering chosen with
$A$ first. The nonlocal SRE $M_\alpha^{\NL}$ is defined in
Eq.~\eqref{eq:m-MNL}, with $M_\alpha$ the stabilizer R\'enyi entropy
\cite{Leone2022SRE}. For fermions, local parity superselection restricts
$U_A,U_B$ to unitaries commuting with the respective local parities
\cite{WWW1952,Banuls2007Fermionic}; this gives $M_\alpha^{f,\NL}$
and the parity-resolved fidelity $\Phi_S^{\NL}$. These local unitaries need not be Gaussian.

The graded Schmidt spectrum is invariant under the parity-even local group.
The fidelity problem of Sec.~\ref{sec:exact} maximizes the overlap with
stabilizer targets of allowed flat spectra; the SRE problem in
Sec.~\ref{sec:small} minimizes Pauli moments over the entire allowed orbit.
A fidelity bound controls the SRE but need not determine its exact minimum.
The corresponding qubit quantities without the
parity constraint are denoted $D_{\min}^{\NL}$
and $M_\alpha^{\NL}$. Logarithms $\log_2$ are in bits,
whereas $\ln$ denotes the natural logarithm. Table~\ref{tab:notation}
collects the notation. 

\paragraph*{Gaussian-restricted nonlocal SRE.}
For BCS occupations $p_j$, $q_j=1-p_j$, and $x_j=(1-2p_j)^2$, the
Gaussian-restricted expression studied in
Refs.~\cite{Iannotti2026FNL,Collura2026NLFF} is
\begin{equation}
 \mathfrak M_\alpha(p)
 =\sum_j\frac{1}{1-\alpha}
   \log_2\!\left[\frac{1+x_j^\alpha+(1-x_j)^\alpha}{2}\right].
 \label{eq:Mfree}
\end{equation}
Its identification with the minimum over all local Gaussian unitaries is
proved at $\alpha=2$ for arbitrary subsystem size in
Ref.~\cite{Iannotti2026FNL}; the corresponding statement for $\alpha>2$
is conjectural there. At $\alpha=2$, using $1-x_j=4p_jq_j$,
Eq.~\eqref{eq:Mfree} reduces to
$\mathfrak M_2(p)$ in Eq.~\eqref{eq:m-Mfree2}
of the main text.
For occupations bounded away from $0$, $\frac12$, and $1$, this
Gaussian-restricted quantity is extensive in the number of BCS pairs.

\begin{table}[!t]
\caption{Notation.}
\label{tab:notation}
\footnotesize
\begin{ruledtabular}
\begin{tabular}{p{0.34\columnwidth}p{0.58\columnwidth}}
$m$; $p_j,q_j=1-p_j$ & Number of BCS pairs and occupations (Sec.~\ref{app:exact})\\
$\lambda^{(\pm)}$; $a_i^{(\pm)}$ & Graded probabilities and sorted
amplitudes (Sec.~\ref{app:exact}; Eq.~\eqref{eq:PhiS})\\
$\lambda_i$; $\mu_s$ &
Ungraded Schmidt probabilities;
$\mu_s=\sqrt{\lambda_s}$ denotes the Schmidt
amplitudes in the proof of
Proposition~\ref{thm:small-gauss}\\
$\Phi^{\rm q}=2^{-D_{\min}^{\NL}}$ & Qubit fidelity
(Sec.~\ref{app:exact}; Eq.~\eqref{eq:sscost})\\
$\Phi_S^{\NL}=2^{-\Dm}$ & Parity-resolved fidelity. $S$ stands for superselection [Eq.~\eqref{eq:PhiS}]\\
$\Phi_*$; $b$ & Relaxed flat-spectrum fidelity and its kernel
[Eq.~\eqref{eq:kernel}]\\
$M_\alpha$; $W=2^{-M_2}$ & SRE; stabilizer purity
[Eq.~\eqref{eq:s3b-purity}]\\
$M_\alpha^{\NL}$; $M_\alpha^{f,\NL}$ & Qubit and parity-even nonlocal
SRE [Eq.~\eqref{eq:m-MNL}]\\
$\mathfrak M_\alpha(p)$ & non-local SRE over local Gaussian unitaries alone
[Eq.~\eqref{eq:m-Mfree2}]\\
$\phi_a$ & Graded reference state [Eq.~\eqref{eq:ref}]\\
$X$; $C_E$ & Entanglement energy and capacity [Eq.~\eqref{eq:m-CE}]\\
$c_*$ & Pair-capacity maximum [Eq.~\eqref{eq:m-cstar}]\\
$\Phi_S^{U(1)}$ & Number-conserving fidelity with definite-number
targets [Eq.~\eqref{eq:u1prod}; Proposition~\ref{prop:u1}]\\
\end{tabular}
\end{ruledtabular}
\end{table}

\paragraph*{Relation to prior work.}
For qubits, the exact nonlocal stabilizer fidelity
is determined by the Schmidt spectrum
\cite{Sierant2026Exact,Viscardi2026Purification},
while dyadic spectral bounds control the nonlocal
SRE up to additive constants
\cite{Sierant2026Spectral}.
Capacity-based bounds and smooth-spectrum
asymptotics were studied in
Ref.~\cite{Aditya2026Spreading}.
The relationship between spectrum flatness and
magic was examined in
Refs.~\cite{Cao2025Gravitational,
Liu2026AntiFlatness,Tirrito2024Flatness}.
The pure-state inequality $M_2\le4D_{\min}$
\cite{Haug2023Stabilizer,Sierant2026Exact}
is extended to $\alpha>1$ in
Eq.~\eqref{eq:MD}.
The Gaussian-restricted result used here is from
Refs.~\cite{Iannotti2026FNL,Collura2026NLFF}.

\paragraph*{Numerical methods.}
For many BCS pairs, listing all $2^m$ Schmidt levels to evaluate
Eq.~\eqref{eq:m-PhiSdef} is impractical. Write $X=X_{\min}+Y$, where $Y$ is a sum of
independent Bernoulli energy increments with
gaps $g_j$ as defined in
Proposition~\ref{prop:bcs-converse}.
We propagate the distribution of $Y$ with parity resolution on a
$10^{-3}$-bit grid, splitting weights linearly between neighboring
bins. Cumulative probabilities and root sums evaluate the pure and
balanced flat-template fidelities, without enumerating Schmidt levels;
truncation at $Y_{\rm cut}$ is exact for thresholds below it. Against
full enumeration for random occupations at $m=16,20$, the error is at
most $1.5\times10^{-7}$ bits; against the exact weight classes of
2048 identical pairs it is at most $1.3\times10^{-3}$ bits.
The occupations are extracted from block correlation matrices: an
antiperiodic ring of 65536 sites for chains and a $1024\times1024$
torus for the two-dimensional Fermi sea
($\varepsilon_k=-2(\cos k_x+\cos k_y)<\mu_F=-1$, filling $0.31$).
Haar-random Gaussian states are averaged over seeds. The contributions
in Fig.~\ref{fig:m-fidelity}(b) use the kernel of
Eq.~\eqref{eq:kernel}; reference-state SREs use the Pauli fourth-moment
formula [Eq.~\eqref{eq:Wsum}] evaluated with fast Walsh--Hadamard
transforms.

The ground states of the interacting chains
discussed in Sec.~\ref{sec:m-laws-par}
were obtained using DMRG: open boundaries, a cut in the middle, conserved particle number (and $S^z=0$ for Hubbard chains), bond dimension $128$ ($t$--$V$) and $384$ (Hubbard), and truncation errors at most $6\times10^{-4}$. The chains are at half filling; for odd $L$ we take $N=(L+1)/2$, which selects one charge-density-wave pattern for $V>2t$, as does an edge pinning field for even $L$. At $V=0$ and $U=0$ the exact $C_E$ is reproduced to within $0.3\%$. For these states $M_2^{f,\NL}$ is bounded from above by the graded reference state of Eq.~\eqref{eq:ref} and from below by the staircase bound of Ref.~\cite{Sierant2026Spectral}: if $\nu^\uparrow$ repeats $\lambda_{2^{k-1}+1}$ on each dyadic shell $2^{k-1}<R\le2^k$ (and $\nu^\uparrow_1=\lambda_1$), then
\begin{equation}
M_2^{f,\NL}\ge M_2^{\NL}\ge-\log_2W\big(\sqrt{\nu^\uparrow}\big),\label{eq:Qsh2}
\end{equation}
\CHANGE{with $W$ the sum of Eq.~\eqref{eq:Wsum} on the ordered amplitudes $\sqrt{\nu^\uparrow}$. When $W>1$ this bound is negative and we use $M_2^{f,\NL}\ge0$ instead.}

\section{Exact nonlocal stabilizer fidelity}\label{sec:exact}

\subsection{Exact fidelity from graded spectra}\label{app:exact}

Let $\ket{\psi}$ be a pure state of definite
total fermion parity on $N=N_A+N_B$ modes,
and let $n=\min(N_A,N_B)$ be the number of
modes on the smaller side.
Each parity sector contains at most $2^{n-1}$
nonzero Schmidt values.
The reduced state $\rho_A$ commutes with the local parity $P_A$, so its 
entanglement spectrum splits into $\lambda = (\lambda^{(+)}, \lambda^{(-)})$. 
Write $ a_i^{(\eta)} = \sqrt{\lambda_i^{(\eta)\downarrow}}$ for the Schmidt 
amplitudes sorted within each sector $\eta=\pm$.

\paragraph*{Local orbits.} 
Two pure states of the same total parity are
related by parity-even local unitaries if and
only if their graded Schmidt spectra agree. More generally, the largest 
overlap between the local orbits of states $\ket\psi$ and $\phi$ with graded spectra 
$\lambda$ and $\nu$ is 
\begin{equation}
\max_{U_A,U_B}\bigl|\langle\phi|U_A\otimes U_B|\psi\rangle\bigr|
 =\beta(\lambda,\nu)
 :=\sum_{\eta=\pm}\sum_i
 \sqrt{\lambda_i^{(\eta)\downarrow}\nu_i^{(\eta)\downarrow}}.
\label{eq:beta}
\end{equation}

Indeed, the Schmidt vectors can be chosen to have definite local parity. Parity-even unitaries rotate their bases independently in the two sectors, and von Neumann's trace inequality bounds each sector's overlap by the sum of products of the ordered Schmidt amplitudes. Aligning both bases and their relative phases attains the sum in Eq.~\eqref{eq:beta} (the graded counterpart of Ref.~\cite{Sierant2026Exact}).

\paragraph*{Allowed stabilizer spectra.}
For stabilizer targets of definite total parity, the nonzero Schmidt
values are flat, and their parity split must be one of two kinds:
\emph{pure}, with all $2^k$ levels in one sector ($0\le k\le n-1$),
or \emph{balanced}, with $2^{k-1}$ levels in each sector
($	1\le k\le n$). Denote this set of spectra by $\Sf$.

Here is the classification and its converse. For a stabilizer state
$s$, let $\mathsf S_A$ be its stabilizers supported on $A$. Then
$\rho_A=2^{-N_A}\sum_{g\in\mathsf S_A}g$ is proportional to a projector,
and all $g\in\mathsf S_A$ commute with $P_A$. If
$\pm P_A\in\mathsf S_A$, the support lies in one parity sector;
otherwise $\Tr(P_A\rho_A)=0$, so the flat levels occur equally in
both sectors. Conversely, $k$ fermionic Bell pairs realize a balanced
split, whereas
\begin{equation*}
 2^{-k/2}\sum_{s\in\{0,1\}^k}
 \ket{s,\,|s|\bmod2}\ket{s,\,|s|\bmod2}
\end{equation*}
realizes a pure split: the final mode on each side stores the parity
of $k$ Bell pairs. Both are stabilizer states, the latter obtained
by local Clifford gates. Appropriate bit flips realize either sector
and either total parity. Finally, an indefinite-parity stabilizer
target cannot improve the fidelity: if
$P\ket\psi=\pi\ket\psi$ and $\langle s|P|s\rangle=0$, then
$(I+\pi P)\ket s/\sqrt2$ is a parity-definite stabilizer target
whose squared overlap with $\ket\psi$ is twice as large.

\paragraph*{Exact formula.}
The fermionic nonlocal stabilizer fidelity is
$\Phi^{\NL}_S=\max_{U_A,U_B}\max_{s\in\mathrm{STAB}}
|\langle s|U_A\otimes U_B|\psi\rangle|^2$, with parity-even
$U_A,U_B$. Maximizing Eq.~\eqref{eq:beta} over $\nu\in\Sf$ gives
\begin{align}
\Phi^{\NL}_S=\max\Bigg\{&
\max_{\substack{0\le k\le n-1\\\eta=\pm}}
2^{-k}\Big(\sum_{i=1}^{2^k}a_i^{(\eta)}\Big)^2,\nonumber\\
&\max_{	1\le k\le n}2^{-k}
\Big(\sum_{i=1}^{2^{k-1}}[a_i^{(+)}+a_i^{(-)}]\Big)^2\Bigg\}.
\label{eq:PhiS}
\end{align}
The two lines correspond, respectively, to pure and balanced splits.
Thus $\Phi^{\NL}_S=\max_{\nu\in\Sf}\beta(\lambda,\nu)^2$ depends
only on the graded spectrum. Operationally, the targets are Bell pairs
or Bell pairs with a copied parity bit, reached by parity-even local
unitaries without communication. Appending local vacuum modes does
not change Eq.~\eqref{eq:PhiS}: newly admissible templates have more
slots than the old parity-sector dimensions, but no additional
nonzero Schmidt amplitudes. For the same reason the upper limits on $k$
never bind: a sector holds at most $	2^{n-1}$ nonzero values, and a term
with more slots than that is smaller than the term with half as many,
so the maxima may run over all $k$, as in Eq.~\eqref{eq:m-PhiS} of the
main text, and $\Phi^{\NL}_S$ depends neither on $N_A$, $N_B$ nor on
unpaired modes.

For $m$ BCS pairs, the graded Schmidt spectrum
of the Gaussian normal form
\cite{Botero2004BCS} has probabilities
$\prod_j p_j^{s_j}(1-p_j)^{1-s_j}$
and sector $(-1)^{|s|+c}$, where
$c\in\{0,1\}$ accounts for unpaired
occupied modes.

\paragraph*{Two immediate consequences.}
Write $\Phi^{\rm q}=2^{-D_{\min}^{\NL}}$
for the qubit fidelity and $\Dm=-\log_2\Phi^{\NL}_S$. Parity superselection costs
at most one bit:
\begin{equation}
 D_{\min}^{\NL}\le\Dm\le D_{\min}^{\NL}+1.
\label{eq:sscost}
\end{equation}
To see the upper bound, take the $2^k$ levels of an optimal qubit
flat template. If $k<n$, they fit inside a balanced fermionic
template of total rank $2^{k+1}$, with at least half the qubit
fidelity. If $k=n$, the full template is already balanced.

The pure-state SRE--fidelity inequality~\cite{Haug2023Stabilizer,Sierant2026Exact} extends to $\alpha>1$: if $\ket s$ has stabilizer group $G$, the power mean of order $2\alpha$ gives $|\braket s\psi|^2=2^{-N}\sum_{g\in G}\bra\psi g\ket\psi\le\big(2^{-N}\sum_P\bra\psi P\ket\psi^{2\alpha}\big)^{1/(2\alpha)}=2^{-(\alpha-1)M_\alpha/(2\alpha)}$, so $M_\alpha\le\frac{2\alpha}{\alpha-1}D_{\min}$. Hence, for $\alpha>1$,
\begin{equation}
\begin{aligned}
 M_\alpha^{f,\NL}(\psi)&\le\frac{2\alpha}{\alpha-1}\Dm,\\
 M_2^{f,\NL}(\psi)&\le4\Dm\qquad(\alpha=2).
\end{aligned}
\label{eq:MD}
\end{equation}
Apply the inequality to a parity-even local
unitary attaining $\Phi_S^{\NL}$. This bound also supplies the $T$-count
interpretation discussed in the main text, using the single-gate
bound of Ref.~\cite{Zhu2024Amortized}.

\subsection{A universal bound from the capacity of entanglement}
\label{app:capacity}

Let $\lambda_1\ge\lambda_2\ge\cdots$ be the sorted ungraded
Schmidt probabilities.
Regard $X=-\log_2\lambda$ as a random variable taking
$X_i=-\log_2\lambda_i$ with probability $\lambda_i$. Its variance
is $\operatorname{Var}X=C_E/(\ln2)^2$, where $C_E$ is the capacity of
entanglement in natural units~\cite{Yao2010Capacity,deBoer2019Capacity}.

\paragraph*{The flat-truncation identity.}
Dropping the parity and power-of-two constraints, and defining
$b(z):=\big(2\cdot2^{z/2}-2^z\big)_+$, gives the identity
\begin{equation}
 \Phi_*:=\max_{K\ge1}\frac1K
 \Big(\sum_{i=1}^K\sqrt{\lambda_i}\Big)^2
 =\sup_{y\in\mathbb R}\mathbb E_\lambda\,b(X-y).
\label{eq:kernel}
\end{equation}
To prove the identity, put $a_i=\sqrt{\lambda_i}$. Completing the
square gives, for every nonempty set of indices $T$,
\begin{equation*}
 \frac{(\sum_{i\in T}a_i)^2}{|T|}
 =\max_{t\ge0}\sum_{i\in T}(2ta_i-t^2).
\end{equation*}
At fixed $t$, the maximizing set contains exactly the positive
summands; at fixed size, it contains the largest amplitudes.
Consequently,
\begin{equation*}
 \Phi_*=\max_{t\ge0}\sum_i(2ta_i-t^2)_+.
\end{equation*}
Substituting $t=2^{-y/2}$ turns each term into
$\lambda_i b(X_i-y)$ and proves Eq.~\eqref{eq:kernel}.

\begin{proposition}[Capacity bound]\label{prop:capbound}
Every pure state of definite fermion parity obeys
\begin{equation}
\begin{aligned}
 \Phi_*&\ge\frac{2}{\sqrt{3(C_E+5)}}, \quad \Phi^{\rm q}\ge\frac{\Phi_*}{\sqrt2},\\
 \Phi_S^{\NL}&\ge(\sqrt2-1)\Phi_* \ge\frac{2(\sqrt2-1)}{\sqrt{3(C_E+5)}}.
\end{aligned}
\label{eq:capbound}
\end{equation}
In particular,
\begin{equation*}
\begin{aligned}
 D_{\min}^{\NL}&\le\tfrac12\log_2\!\left[\tfrac32(C_E+5)\right],\\
 \Dm&\le\tfrac12\log_2\!\left[
 \tfrac34(3+2\sqrt2)(C_E+5)\right].
\end{aligned}
\end{equation*}
Balanced targets alone, i.e., fermionic Bell-pair conversion,
achieve fidelity at least $[6(C_E+5)]^{-1/2}$.
The bounds on $\Phi_*$ and $\Phi^{\rm q}$ also hold for qubits.
\end{proposition}

\begin{proof}
\emph{Smoothing and the peak bound.}
Set $\gamma=\ln2$. The function
\begin{equation*}
 g(t):=\frac\gamma4 b(2-t)
 =\gamma\big(e^{-\gamma t/2}-e^{-\gamma t}\big),\qquad t\ge0,
\end{equation*}
(and $g(t)=0$ for $t<0$) is the density of $E_1+E_2$,
where $E_1,E_2$ are independent exponential random variables of
rates $\gamma/2$ and $\gamma$. With $V=2-E_1-E_2$ independent
of $X$, Eq.~\eqref{eq:kernel} reads
\begin{equation*}
 \Phi_*=\frac4\gamma\,\|\rho_{X-V}\|_\infty,
 \qquad \operatorname{Var}(X-V)=\frac{C_E+5}{\gamma^2}.
\end{equation*}
For any probability density $\rho$ bounded by $M$, its variance
is at least $1/(12M^2)$: among densities bounded by $M$, the
uniform density of height $M$ on an interval minimizes the
second moment about any fixed center. Taking the center to be
the mean yields $\|\rho\|_\infty\ge(12\operatorname{Var}\rho)^{-1/2}$.
Applying this to $\rho_{X-V}$ gives the first inequality in
Eq.~\eqref{eq:capbound}.

\emph{Rounding to a power of two.}
For any nonnegative decreasing sequence $z_i$, let
$f(h)=(\sum_{i=1}^h z_i)^2/h$. If $L\le h\le U$ with
$L,U$ consecutive powers of two, monotonicity implies
$f(L)\ge(L/h)f(h)$ and $f(U)\ge(h/U)f(h)$.
Hence
\begin{equation*}
 \max\{f(L),f(U)\}\ge\frac{f(h)}{\sqrt2}
 \qquad(L\le h\le U).
\end{equation*}
Applying this to the ungraded amplitudes proves
$\Phi^{\rm q}\ge\Phi_*/\sqrt2$.

\emph{Rounding with parity.}
Put $R=2^{n-1}$ and
$S_\eta(h)=\sum_{i=1}^{h}a_i^{(\eta)}$.
Let $F_{\rm gr}$ be the maximum of
\[
\frac{S_+(h)^2}{h},\qquad
\frac{S_-(h)^2}{h},\qquad
\frac{[S_+(h)+S_-(h)]^2}{2h},
\]
over integers $1\le h\le R$.

Choose a maximizing truncation for $\Phi_*$,
containing $h$ levels of one parity and
$\ell\le h$ of the other, with amplitude
sums $A$ and $B$. Then
\[
\Phi_*=\frac{(A+B)^2}{h+\ell},
\qquad h\le R.
\]
If $\ell=0$, then $F_{\rm gr}\ge\Phi_*$.
For $\ell>0$, the pure and balanced
candidates give
\[
F_{\rm gr}\ge
\max\left\{
\frac{A^2}{h},\frac{B^2}{\ell},
\frac{(A+B)^2}{2h}
\right\}.
\]
Since
\[
\max\left\{\frac{A^2}{h},\frac{B^2}{\ell}\right\}
\ge
\frac{(A+B)^2}{(\sqrt h+\sqrt\ell)^2},
\]
setting $x=\ell/h$ yields
\[
\frac{F_{\rm gr}}{\Phi_*}
\ge
\max\left\{
\frac{1+x}{(1+\sqrt x)^2},
\frac{1+x}{2}
\right\}
\ge2-\sqrt2.
\]
The two expressions cross at
$x=(\sqrt2-1)^2$.
Applying the preceding rounding estimate
to $a_i^{(+)}$, $a_i^{(-)}$, and
$a_i^{(+)}+a_i^{(-)}$ gives
\[
\Phi_S^{\NL}
\ge\frac{F_{\rm gr}}{\sqrt2}
\ge(\sqrt2-1)\Phi_*.
\]

This proves Eq.~\eqref{eq:capbound}. Finally, the balanced
template containing the $2^k$ largest ungraded levels has at
least half the corresponding qubit fidelity (and agrees with
it when $k=n$). Therefore the best balanced target achieves
$\Phi_S^{\mathrm{bal}}\ge\Phi^{\rm q}/2
\ge[6(C_E+5)]^{-1/2}$, as claimed.
\end{proof}

The capacity bound holds at every system size; it gives no
converse inequality from $C_E$ alone. A single Schmidt level
of weight close to one can keep $\Phi_S^{\NL}$ close to one
even when the remaining levels make $C_E$ arbitrarily large.
For BCS states, $C_E\le c_*m$ [Eq.~\eqref{eq:m-CEbcs}].

The bound of Proposition~\ref{prop:capbound} improves
the capacity-based estimate of
Ref.~\cite{Aditya2026Spreading}, including the
one-bit parity-superselection correction in
Eq.~\eqref{eq:sscost}.
The first inequality of Eq.~\eqref{eq:capbound}
is asymptotically sharp: for spectra
$\lambda_R\propto R^{-1}$ supported on
$1\le R\le 2^L$, the ratio of its two sides
tends to one as $L\to\infty$.

\subsection{Bounds from the number of modes and from the spread of $X$}
\label{app:finite-fidelity}

We complement the capacity bound with two finite-size estimates: a
lower bound in terms of the number of local modes, valid for every
pure state, and a converse for BCS spectra. Together they establish
the optimal $m^{-1/2}$ fidelity scaling when all BCS occupations stay
uniformly away from $0$, $\frac12$, and $1$.

\begin{proposition}[Dimension-dependent fidelity bound]
\label{prop:dyadic-fidelity}
For a pure state of definite total fermion parity with $n\ge2$
modes on the smaller side,
\[
 \Phi_S^{\mathrm{NL}}\ge\frac{3(2+\sqrt2)}{8n},
 \qquad
 D_{\min}^{f,\mathrm{NL}}
 \le\log_2n-\log_2\!\frac{3(2+\sqrt2)}8.
\]
\end{proposition}

\begin{proof}
Let $\lambda_1\ge\cdots\ge\lambda_{2^M}\ge0$ have total
weight $w$, where $M\ge1$, and define
$F_k=2^{-k}(\sum_{i\le2^k}\sqrt{\lambda_i})^2$.
Monotonicity gives
$F_k\ge2^{-k}\sum_{j\le2^k}(2j-1)\lambda_j$.
The sum of these linear minorants is minimized over decreasing
spectra of total weight $w$ at an initial flat spectrum:
$\lambda_1=\cdots=\lambda_h=w/h$, with all later entries zero.
Indeed, every decreasing spectrum is a convex combination of such
vectors. For $2^K\le h<2^{K+1}$, putting $t=h/2^K$ gives the sum
$w(\alpha/t+\beta t)$, where
$\alpha=2-2^{-K}$ and $\beta=1-2^{K-M}$.
At $h=1$ and $h=2^M$ this sum equals
$w(2-2^{-M})\ge\frac32 w$; for the other values,
$K\ge1$, $\alpha\ge\frac32$, $\beta\ge\frac12$, and
$\alpha/t+\beta t\ge2\sqrt{\alpha\beta}\ge\sqrt3$.
Consequently,
\begin{equation}
 \sum_{k=0}^{M}F_k\ge\frac32w.
 \label{eq:dyadic-one-sector}
\end{equation}

Apply this to the two parity sectors with $M=n-1$.
Write $x_j^2,y_j^2$ for their pure-template fidelities
at rank $2^j$. The balanced template of total rank $2^{j+1}$
has fidelity $(x_j+y_j)^2/2$, and
$\max\{x^2,y^2,(x+y)^2/2\}
\ge(2+\sqrt2)(x^2+y^2)/4$ for $x,y\ge0$.
Since $\sum_{j=0}^{n-1}(x_j^2+y_j^2)\ge\frac32$,
one of the $n$ admissible ranks attains the stated bound.
\end{proof}

For the converse, let $X=-\log_2\lambda$ under the ungraded
Schmidt distribution and set
$Q_L=\sup_x\Pr(x-L<X\le x)$ for $L>0$.

\begin{proposition}[Converse for BCS spectra]
\label{prop:bcs-converse}
For every pure state and $L>0$,
\begin{equation}
 \Phi_S^{\mathrm{NL}}\le\Phi^{\rm q}\le\Phi_*
 \le\frac{2Q_L}{1-2^{-L/2}}.
 \label{eq:window-converse}
\end{equation}
For $m$ BCS pairs with $C_E>0$, omit deterministic pairs with
$p_j\in\{0,1\}$ and set
$\bar p_j=\min(p_j,1-p_j)$,
$g_j=\log_2[(1-\bar p_j)/\bar p_j]$, and
$s^2=\operatorname{Var}X=C_E/(\ln2)^2$.
Define the nonnegative quantity
\(
 r=s^{-2}\sum_j\bar p_j(1-\bar p_j)
 [\bar p_j^2+(1-\bar p_j)^2]g_j^3
 \le\max_j g_j.
\)
Then, with a Berry--Esseen constant $C_0\le0.56$,
\begin{equation}
 \begin{aligned}
 \Phi_S^{\mathrm{NL}}&\le\frac{2^{a_{\rm BE}(r)}}{\sqrt{C_E}},\\
 a_{\rm BE}(r)&=\inf_{L>0}\log_2\!\left[
 \frac{2\ln2\,\big(L/\sqrt{2\pi}+2C_0r\big)}
 {1-2^{-L/2}}\right].
 \end{aligned}
 \label{eq:BE-converse}
\end{equation}
If $\bar p_j\in[\delta,\frac12-\delta]$ for all $j$,
$0<\delta<\frac14$, then $C_E\ge c_\delta m$, where
$c_\delta=\min\{c(\delta),c(\frac12-\delta)\}>0$ and
$c(p)=p(1-p)\ln^2[(1-p)/p]$.
Consequently, $\Phi_S^{\mathrm{NL}}=O(m^{-1/2})$;
for $p_j=\frac14$, one obtains
$r=\frac58\log_2 3$, $a_{\rm BE}(r)\simeq2.30$, and
$\Phi_S^{\mathrm{NL}}\le10.4/\sqrt m$.
\end{proposition}

\begin{proof}
Choose $K$ attaining $\Phi_*=A^2/K$, with
$A=\sum_{i\le K}\sqrt{\lambda_i}$ and
$x=-\log_2\lambda_K$.
The variational characterization in Eq.~\eqref{eq:kernel}
implies that the optimal threshold $t=A/K$ satisfies
$t\le2\sqrt{\lambda_K}=2^{1-x/2}$.
Moreover,
$A=\sum_{i\le K}\lambda_i2^{X_i/2}\le\mathbb E_\lambda[2^{X/2}\mathbf1_{X\le x}]$.
Partition $(-\infty,x]$ into intervals of length $L$.
Each carries probability at most $Q_L$, hence
$A\le2^{x/2}Q_L/(1-2^{-L/2})$.
Multiplication by $t$ proves Eq.~\eqref{eq:window-converse}.

For BCS pairs, $X$ is a constant plus independent Bernoulli
increments of sizes $g_j$. Their variance is $s^2$ and the sum
of their third absolute central moments is $rs^2$.
Berry--Esseen bounds the distance between their distribution
function and the normal distribution with the same mean and
variance by $C_0r/s$. It follows that
$Q_L\le[L/\sqrt{2\pi}+2C_0r]/s$.
Use this in Eq.~\eqref{eq:window-converse} and optimize over $L$.
The occupation condition bounds $r$ and ensures
$C_E\ge c_\delta m$, giving the stated scaling.
\end{proof}

If $C_E=0$, every BCS pair is empty, filled, or maximally
entangled, so the state is a stabilizer state and
$\Phi_S^{\mathrm{NL}}=1$.
Together with Proposition~\ref{prop:capbound} and
$C_E\le c_*m$, the converse establishes $\Phi_S^{\mathrm{NL}}=\Theta(m^{-1/2})$
under the stated uniform occupation condition.

\section{Small sizes: when local Gaussian unitaries are optimal}\label{sec:small}
\subsection{Graded reference states}\label{app:small}

Let $\lambda=(\lambda^{(+)},\lambda^{(-)})$
be the graded Schmidt spectrum of a pure
even-parity state $\ket{\psi}$.
After padding the two sides with local vacuum
modes if necessary, let $n$ be the number
of modes on each side. A \emph{graded
labeling} $a$ assigns distinct Fock strings $a_i\in\{0,1\}^n$ to
its Schmidt levels, preserving the parity of each level. The associated
\emph{graded reference state} is
\begin{equation}
 \ket{\phi_a}=\sum_i\sqrt{\lambda_i}\,
 \ket{a_i}_A\ket{a_i}_B .\label{eq:ref}
\end{equation}
Its graded spectrum agrees with that of $\ket\psi$. By the graded Schmidt
orbit characterization in Sec.~\ref{app:exact}, $\phi_a$ is therefore
reachable by parity-even local unitaries, and
$M_\alpha^{f,\NL}(\ket\psi)\le M_\alpha(\phi_a)$. For odd total parity,
flip a fixed bit of every $B$ label; the same conclusion holds.
Thus optimizing over graded labelings supplies explicit trial states.

For $m$ BCS pairs, the \emph{product labeling} $a=s$ uses the
pair-occupation string $s$. In $A$-first Jordan--Wigner order, the
physical BCS state and $\phi_s$ differ by the parity-even diagonal
Clifford $\prod_{i<j}\mathrm{CZ}_{A_iA_j}$, which removes the phase
$(-1)^{|s|(|s|-1)/2}$. Consequently, at the Rényi index $\alpha=2$ used here,
$M_2(\phi_s)=\mathfrak M_2(p)$. In particular, any graded
labeling with $M_2(\phi_a)<\mathfrak M_2(p)$ witnesses failure of
Gaussian optimality.

\begin{figure*}[t]
    \centering
    \resizebox{0.85\textwidth}{!}{%
        \begin{tikzpicture}[
  font=\footnotesize, text=SBink,
  box/.style={rounded corners=5pt, align=flush center, text width=#1,
    inner xsep=6pt, inner ysep=5pt, fill=white, draw=SBline, line width=0.5pt},
  card/.style={box=#1, font=\scriptsize, inner xsep=4pt, inner ysep=4pt},
  lift/.style={blur shadow={shadow blur steps=7, shadow xshift=0pt,
    shadow yshift=-1.2pt, shadow opacity=18}},
  bdgM/.style={circle, inner sep=0pt, minimum size=5.4mm, font=\footnotesize\bfseries,
    fill=SBblue, text=white, draw=white, line width=0.6pt},
  bdgX/.style={circle, inner sep=0pt, minimum size=6.6mm, font=\small\bfseries,
    fill=SBred, text=white, draw=white, line width=0.8pt},
  conn/.style={draw=SBblue, line width=0.85pt, -{Latex[length=2.3mm,width=1.9mm]}},
  thin/.style={draw=SBblue!70, line width=0.6pt, -{Latex[length=1.7mm,width=1.4mm]}},
  stage/.style={circle, minimum size=5.4mm, inner sep=0pt,
    fill=SBblue, text=white, font=\footnotesize\bfseries},
  stlab/.style={rotate=90, fill=SBpanel, inner xsep=2pt, inner ysep=0.5pt,
    font=\scriptsize\scshape, text=SBblue},
]
\definecolor{SBpanel}{HTML}{F4F7F9}\definecolor{SBpanelline}{HTML}{D3DCE3}
\definecolor{SBband}{HTML}{E6EDF2}\definecolor{SBline}{HTML}{B9C6D1}
\definecolor{SBink}{HTML}{22313D}\definecolor{SBmute}{HTML}{5A6B79}
\definecolor{SBblue}{HTML}{4B6A89}\definecolor{SBred}{HTML}{AD4F48}
\definecolor{SBcream}{HTML}{FCF9F2}\definecolor{SBsand}{HTML}{F3E9D6}
\definecolor{SBsplit}{HTML}{E0EAF3}
\def\xc{0.4}   
\def\SBn#1{{\color{SBblue}\bfseries#1}\hspace{0.45em}}   
\def\SBfact#1{\parbox[t]{3.3cm}{\centering\scriptsize#1}}  
\def\SBsep{\hspace{2mm}{\color{SBred!45}\vrule width 0.5pt}\hspace{2mm}}

\node[box=14.2cm, lift, inner ysep=7pt, anchor=north,
      path picture={\fill[SBred] (path picture bounding box.north west)
        rectangle ([yshift=-2.4pt]path picture bounding box.north east);}]
  (TH) at (\xc,-0.35) {%
  {\large\bfseries Proposition~\ref{thm:small-gauss}.}\ {\normalsize For $m\le 3$ BCS pairs,}\\[3pt]
  {\normalsize$M_2^{f,\mathrm{NL}}(\Psi_m)=\mathfrak M_2(p)
    =-\sum_{j=1}^{m}\log_2\!\big[1-4p_jq_j(1-2p_j)^2\big]$}\\[3pt]
  {\small\itshape the BCS state minimizes $M_2$ over its
parity-even local-unitary orbit}};

\foreach \i/\x in {1/-5.11, 2/-1.47, 3/2.17, 4/5.81} \coordinate (c\i) at (\x,0);
\tikzset{acard/.style={card=3.12cm, anchor=north, draw=none, fill=none}}
\node[acard] (A1) at ($(c1 |- TH.south)+(0,-0.55)$)
  {{\footnotesize\SBn1\bfseries Reduce to $m=3$}\\[2pt]
   append vacuum pair(s), i.e.\ set added $p_j=0$; vacuum factors leave the
   optimized $M_2$ unchanged};
\node[acard] (A2) at ($(c2 |- TH.south)+(0,-0.55)$)
  {{\footnotesize\SBn2\bfseries Reference state}\\[2pt]
   three-pair BCS state $\to$ product-labeled $\phi_M$; same $M_2$ up to a
   parity-even diagonal Clifford};
\node[acard] (A3) at ($(c3 |- TH.south)+(0,-0.55)$)
  {{\footnotesize\SBn3\bfseries Orbit \smash{$N=UMU^\dagger$}}\\[2pt]
   optimizing over parity-even local unitaries reduces to this one-unitary
   orbit};
\node[acard] (A4) at ($(c4 |- TH.south)+(0,-0.55)$)
  {{\footnotesize\SBn4\bfseries Graded blocks}\\[2pt]
   $N=A\oplus B$ on $\mathcal H_+\oplus\mathcal H_-$, each block of rank~$4$};
\path let \p1=(A1.south), \p2=(A2.south), \p3=(A3.south), \p4=(A4.south) in
  coordinate (Abot) at (0,{min(\y1,\y2,\y3,\y4)-1pt});
\coordinate (Amid) at ($(A1.north)!0.5!(A1.north |- Abot)$);
\foreach \i/\j in {1/2,2/3,3/4} \draw[thin] (A\i.east |- Amid) -- (A\j.west |- Amid);
\node[fit=(A1.north west)(A4.east |- Abot), inner xsep=0.2cm, inner ysep=0.2cm] (BA) {};

\coordinate (S5top) at ($(\xc,0 |- BA.south)+(0,-0.45)$);
\node[box=8.5cm, lift, fill=SBsplit, draw=SBblue, line width=0.9pt, inner ysep=6pt,
      anchor=north] (S5) at (S5top) {%
  {\small\bfseries Parity split}\\[2pt]
  Ref.~\cite{Sierant2026Spectral}: write $2^{-M_2}$ as a fixed spectral term plus two
  orbit-dependent terms, $Q_{(1^4)}(N)$ and $c_F(N^2)$\\[1pt]
  split by parity bidegree in $\Lambda^4(\mathcal H_+\oplus\mathcal H_-)$};
\node[bdgX] at ($(S5.north west)+(0.36,0)$) {5};
\draw[conn] (BA.south -| S5.north) -- (S5.north);
\node[box=2.25cm, font=\scriptsize\itshape, fill=SBpanel, draw=SBline,
      dash pattern=on 1.5pt off 1.5pt, inner xsep=4pt, inner ysep=3pt, anchor=west]
  (NT) at ($(S5.east)+(0.28,0)$) {$(3,1)$ and $(1,3)$ vanish by parity selection};
\draw[SBline, line width=0.6pt, dash pattern=on 1pt off 1.5pt] (S5.east) -- (NT.west);

\coordinate (MXc) at (2.52,0);
\node[font=\small\bfseries, anchor=north west] (MXh)
  at ($(-2.7,0 |- S5.south)+(0.55,-0.78)$) {Mixed part {\normalfont\footnotesize$(2,2)$}};
\node[font=\scriptsize\itshape, text=SBmute, anchor=east] at (7.5,0 |- MXh.east)
  {three lemmas};
\foreach \i/\x/\n/\t in {
    1/-0.96/lem:s3b-projector/{graded projector\\ decomposition},
    2/2.52/lem:s3b-diag/{diagonal rank-four\\ estimate},
    3/6.0/lem:s3b-4th/{fourth-moment\\ estimate}} {
  \node[card=2.7cm, minimum height=10.5mm, anchor=north] (L\i)
    at ($(\x,0 |- MXh.south)+(0,-0.12)$)
    {{\footnotesize\bfseries Lemma~\ref{\n}}\\[1pt] \t};
}
\draw[thin] (L1.east) -- (L2.west);
\draw[thin] (L2.east) -- (L3.west);
\node[fit=(MXh)(L1)(L3), inner xsep=0.25cm, inner ysep=0.22cm] (MX) {};
\node[bdgM] at (MX.north west) {6};

\node[box=3.43cm, lift, fill=SBcream, inner ysep=6pt, anchor=north] (PU)
  at (-5.025,0 |- MX.north) {%
  {\small\bfseries Pure part}\\
  {\footnotesize$(4,0)\oplus(0,4)$}\\[3pt]
  {\scriptsize controlled by the four-dimensional compression estimate}\\[2pt]
  {\scriptsize\itshape\color{SBmute} the easier branch}};

\draw[conn] (S5.west) to[out=180, in=90] (PU.north);
\draw[conn] (S5.south -| MXc) -- (MX.north -| MXc);

\path let \p1=(PU.south), \p2=(MX.south) in
  coordinate (Dtop) at (\xc,{min(\y1,\y2)-0.5cm});
\node[box=7.6cm, lift, draw=SBblue, line width=0.7pt, inner ysep=5pt, anchor=north]
  (S7) at (Dtop) {%
  {\small\bfseries Recombine}\\[1pt]
  combine pure and mixed bounds $\Rightarrow$ bounds on $c_F(N^2)$ and $Q_{(1^4)}(N)$};
\node[bdgM] at ($(S7.north west)+(0.32,0)$) {7};
\draw[conn] (PU.south) to[out=-90, in=180] (S7.west);
\draw[conn] (MX.south -| MXc) -- (S7.north -| MXc);

\node[box=11.6cm, lift, fill=SBsand, draw=SBred, line width=0.9pt, inner ysep=7pt,
      anchor=north] (S8) at ($(S7.south)+(0,-0.4)$) {%
  {\small\bfseries\color{SBred} Destination}\\[2pt]
  {\large$M_2^{f,\mathrm{NL}}(\Psi_3)=\mathfrak M_2(p)$}\\[3pt]
  \SBfact{insert into the formula\\ for $2^{-M_2}$}\SBsep
  \SBfact{recover exactly the\\ canonical BCS value}\SBsep
  \SBfact{boundary values\\ by continuity}};
\node[bdgX] at ($(S8.north west)+(0.4,0)$) {8};
\draw[conn] (S7.south) -- (S8.north);

\def\SBstage#1#2#3#4{
  \path let \p1=(#3.north), \p2=(#4.south) in
    coordinate (st) at (-7.45,\y1) coordinate (sb) at (-7.45,\y2);
  \draw[SBblue!45, line width=0.7pt] (st) -- (sb);
  \node[stage] at (st) {#1};
  \node[stlab] at ($(st)!0.55!(sb)$) {#2};}
\SBstage{A}{setup}{BA}{BA}
\SBstage{B}{split}{S5}{S5}
\node[fit=(MX)(PU), inner sep=0pt] (SC) {};
\SBstage{C}{analysis}{SC}{SC}
\SBstage{D}{conclusion}{S7}{S8}

\begin{scope}[on background layer]
  \path let \p1=(S8.south) in
    [fill=SBpanel, draw=SBpanelline, line width=0.6pt, rounded corners=10pt]
    (-8,0) rectangle (8,\y1-0.4cm);
  \fill[SBband, rounded corners=8pt] (BA.south west) rectangle (BA.north east);
  \foreach \i in {1,...,4}
    \filldraw[fill=white, draw=SBline, line width=0.5pt, rounded corners=5pt]
      (A\i.north west) rectangle (A\i.east |- Abot);
  \fill[SBband, rounded corners=8pt] (MX.south west) rectangle (MX.north east);
\end{scope}
\end{tikzpicture}%
    }
    \caption{Proof roadmap for Proposition~\ref{thm:small-gauss}.}
    \label{fig:m3-roadmap}
\end{figure*}
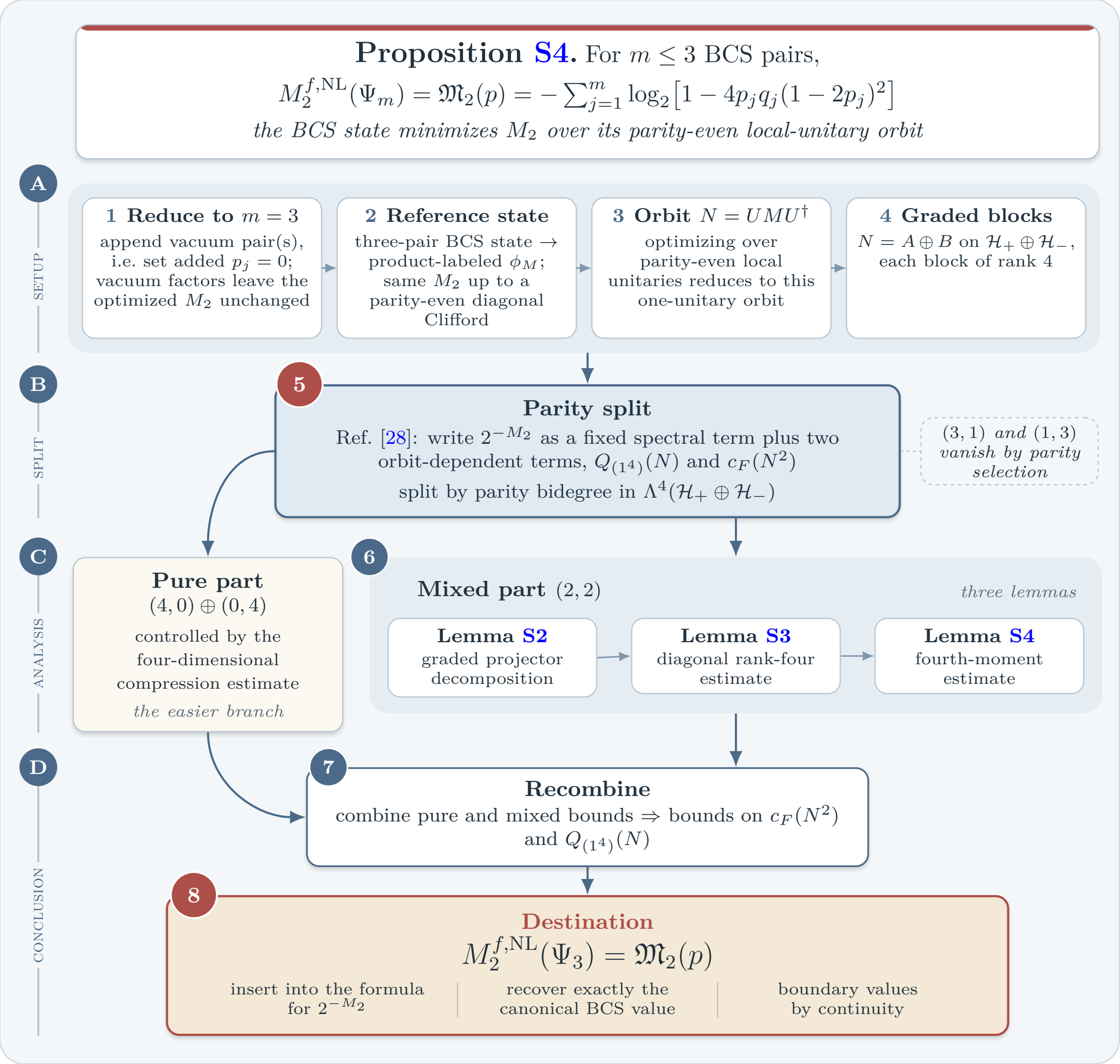

\subsection{Gaussian optimality for up to three BCS pairs}
We \CHANGE{prove} that the Gaussian minimum of Eq.~\eqref{eq:m-Mfree2} is also the minimum over parity-even local unitaries for at most three BCS pairs.

\begin{proposition}[Gaussian optimality for up to three pairs]\label{thm:small-gauss}
For $m \le 3$, let $\Psi_m$ consist of $m$ BCS pairs with occupations $p_j\in [0,1]$, $q_j = 1- p_j$, and any additional local modes in the vacuum. Then
\begin{equation}
\begin{aligned}
  M_2^{f, \mathrm {NL}}(\Psi_m)&= \mathfrak M_2(p)\\
  &=-\sum_j\log_2\big[1-4p_jq_j(1-2p_j)^2\big]
  \label{eq:s3b-main}
\end{aligned}
\end{equation}
\end{proposition}
The minimum is attained by the BCS state $\Psi_m$ itself.

\paragraph*{Reduction to three pairs.}
For $m<3$, append $3-m$ vacuum pairs with occupations
$p_j=0$. Every output of an allowed parity-even
local unitary on the original system embeds, by
tensoring these vacuum modes, into the local orbit
covered by the three-pair statement. Since tensoring
with a stabilizer vacuum leaves $M_2$ unchanged, the
three-pair bound implies
$M_2^{f,\NL}(\Psi_m)\ge\mathfrak M_2(p)$.
The canonical BCS representative attains equality.

\paragraph*{Proof ingredients.}
We use the one-unitary reduction
\cite{Liu2026AntiFlatness,Sierant2026Spectral},
the Pauli-projector identity, and the
compression estimate(Theorem~22 of Ref.~\cite{Sierant2026Spectral}).
Lemmas~\ref{lem:s3b-parity}--\ref{lem:s3b-4th}
establish the parity decomposition and the
rank-four estimates used below.
Figure~\ref{fig:m3-roadmap}
summarizes the proof.

\paragraph*{Setting.}
We treat first the interior case $0<p_j<1$ for all $j$ and then extend to the boundary by continuity. Let $n\ge 3 $ be the number of local modes on each side after padding with vacuum modes if needed, and let $ d = 2^n$, $D= d/ 2$, and \CHANGE{$\mathsf q=4/d=2/D$; $\mathsf q$ is the constant $q$ of Ref.~\cite{Sierant2026Spectral}, not an occupation}. The local parity is $ P =Z^{\otimes n}$, and the parity-even local unitary group is $G=\{U\in U(d):[U,P]=0\}\cong U(D)\times U(D)$, block diagonal in the two parity sectors. \CHANGE{Let $E,O\subset\mathbb F_2^n$ be the Fock strings of even and odd weight, read as integers with the last mode as the least significant bit; this order fixes the signs of the sorted wedges below. The eigenspaces of $P$ with eigenvalues $\pm1$ are $\mathcal H_+=\mathbb C^E$ and $\mathcal H_-=\mathbb C^O$.}

Choose the BCS product labeling by the pair-occupation strings $s\in \mathbb F_2^3$, and define
\begin{equation}
    \lambda_s=\prod_{j=1}^{3}q_j^{1-s_j}p_j^{s_j}
    \qquad
    \mu_s=\sqrt{\lambda_s} .
\end{equation}

Let $M$ be the diagonal $d\times d$ matrix
with entries $\mu_s$ on the eight Fock
configurations of the three paired modes,
with all spectator modes empty, and zero
elsewhere. Let
\begin{equation}
    \phi_M=\sum_{x,y}(M)_{xy}\,|x\rangle|y\rangle
\end{equation}
be the corresponding product-labeled reference state. In the $A$-first Jordan--Wigner order, the physical BCS state differs from $\phi_M$ by the parity-even diagonal Clifford $\prod_{1\le i<j \le 3}\mathrm{CZ}_{A_iA_j}$ on $A$. Thus they have the same stabilizer R\'enyi entropy and the same optimized value of $M_2$.
\CHANGE{The state $\phi_M$ is a tensor product of three single-pair states, so its stabilizer purity, defined below, is $\prod_j(1-4u_j+16u_j^2)=2^{-\mathfrak M_2(p)}$ with $u_j=p_jq_j$, by Eq.~\eqref{eq:m-Mfree2}.}

The one-unitary reduction of Refs.~\cite{Liu2026AntiFlatness,Sierant2026Spectral} applies to the parity-even subgroup $G$ because $G$ is closed under complex conjugation. \CHANGE{Its proof (Lemma~1 of Ref.~\cite{Sierant2026Spectral}) is a Cauchy--Schwarz inequality for the Pauli projector, which bounds the stabilizer purity of $(U\otimes V)\phi_M$ by the geometric mean of those of $(U\otimes\overline U)\phi_M$ and $(\overline V\otimes V)\phi_M$. For $U,V\in G$ both pairs on the right lie again in $G\times G$, and $(U\otimes\overline U)\phi_M=\phi_{UMU^\dagger}$.} Hence
\begin{equation}
\min_{U_A,U_B\in G}M_2\big((U_A\otimes U_B)\phi_M\big)
=
\min_{U\in G}M_2(\phi_{UMU^\dagger}).
\label{eq:s3b-oneunit}
\end{equation}
We therefore fix $U\in G$, set $ N=UMU^\dagger $, and write $\phi_N$ for the corresponding bipartite pure state.

Since $U$ preserves parity, $N$ is block diagonal with respect to $\mathcal H_+\oplus\mathcal H_-$: $N=A\oplus B$,
where $A$ and $B$ are positive semidefinite matrices acting on $\mathcal H_+$ and $\mathcal H_-$, respectively. Each has rank four. \CHANGE{The nonzero eigenvalues $\alpha_1,\dots,\alpha_4$ of $A$ are the amplitudes $\mu_s$ with $s$ even, and those of $B$, $\beta_1,\dots,\beta_4$, the amplitudes with $s$ odd.} Define
\begin{equation}
\begin{aligned}
u&=\prod_{a=1}^{4}\alpha_a=\prod_{a=1}^{4}\beta_a
  =\prod_{j=1}^{3}p_jq_j,\\
T&=\sum_{a<b}(\alpha_a\alpha_b)^4
  =\sum_{a<b}(\beta_a\beta_b)^4.
\end{aligned}
\label{eq:s3b-spectralTu}
\end{equation}
The two equalities follow from the three-pair BCS spectrum. In each parity sector the four Schmidt values multiply to $(\prod_jp_jq_j)^2$, and their six pairwise products are $\{p_j^2,q_j^2\}\times\prod_{k\ne j}p_kq_k$ for $j=1,2,3$; the two multisets of products therefore coincide.

We use the stabilizer purity $W$, as listed in
Table~\ref{tab:notation}:
\begin{equation}
W(\phi):=2^{-M_2(\phi)}.
\label{eq:s3b-purity}
\end{equation}
With $\mathcal P_n=\{I,X,Y,Z\}^{\otimes n}$, define
\begin{equation}
 \Pi_W=d^{-2}\sum_{\sigma\in\mathcal P_n}\sigma^{\otimes4},\qquad
 \Pi_F=\left.\Pi_W\right|_{\Lambda^4\mathbb C^d}.
\end{equation}
The antisymmetric subspace is invariant under $\Pi_W$, so $\Pi_F$ is an orthogonal projector. \CHANGE{For an operator $C$ on $\mathbb C^d$ let $\Lambda^k(C)$ be the operator it induces on $\Lambda^k\mathbb C^d$, and put}
\begin{equation}
\begin{aligned}
Q_{(1^4)}(N)&=\|\Pi_F\Lambda^4(N)\Pi_F\|_{\mathrm{HS}}^2,\\
c_F(C)&=\operatorname{Tr}[\Pi_F\Lambda^4(C)],
\end{aligned}
\end{equation}
\CHANGE{and $I(\mu)=7(\sum_s\mu_s^4)^2-6\sum_s\mu_s^8$. Proposition~2 of Ref.~\cite{Sierant2026Spectral} [its Eq.~(42)] is the identity
\begin{equation}
W(\phi_N)=I(\mu)+6d^2Q_{(1^4)}(N)+18d\,c_F(N^2)-\mathcal D(N),
\label{eq:s3b-identity}
\end{equation}
valid for every unitary $U$, with a defect $\mathcal D(N)\ge0$ that vanishes at $N=M$. Dropping the defect gives}
\begin{equation}
W(\phi_N)
\le
I(\mu)+6d^2Q_{(1^4)}(N)+18d\,c_F(N^2).
\label{eq:s3b-sierant}
\end{equation}
\CHANGE{We write $e_F=e_{x_1}\wedge e_{x_2}\wedge e_{x_3}\wedge e_{x_4}$, with $x_1<x_2<x_3<x_4$, for the sorted unit wedges of computational basis vectors, so that $\langle e_F,\Lambda^4(C)e_{F'}\rangle=\det C[F,F']$. A four-set $F$ with $x_1\oplus x_2\oplus x_3\oplus x_4=0$ is an \emph{affine plane}, a coset of the two-dimensional subspace $L=\{0,x_1\oplus x_2,x_1\oplus x_3,x_1\oplus x_4\}$ of $\mathbb F_2^n$. An \emph{$h$-edge} is a pair $\{x,x\oplus h\}$, and $e_{\{x,x\oplus h\}}$ is its wedge with the smaller string first.} We now bound $Q_{(1^4)}(N)$ and $c_F(N^2)$ in terms of $T$ and $u$.

\CHANGE{\paragraph*{Parity grading of $\Pi_F$.} Since $N$ commutes with $P$, $\Lambda^4\mathbb C^d$ splits into the bidegrees $(k,4-k)$, spanned by wedges of $k$ vectors from $\mathcal H_+$ and $4-k$ from $\mathcal H_-$, and $\Lambda^4(N)$ is block diagonal in them. The next two lemmas show that $\Pi_F$ lives on the \emph{pure} bidegrees $(4,0)$ and $(0,4)$ and on the \emph{mixed} bidegree $(2,2)$, and describe both parts.}

\CHANGE{\begin{lemma}[Parity selection]\label{lem:s3b-parity}
If $v_1,\dots,v_4$ are eigenvectors of $P$ and an odd number of them lie in $\mathcal H_-$, then $\Pi_F(v_1\wedge v_2\wedge v_3\wedge v_4)=0$. Hence $\Pi_F$ vanishes on the bidegrees $(3,1)$ and $(1,3)$.
\end{lemma}}

\begin{proof}
Since $P$ is itself a Pauli string, $\sigma\mapsto P\sigma$ permutes $\mathcal P_n$ up to phases whose fourth powers are one, so $P^{\otimes4}\Pi_W=\Pi_W$, and $P^{\otimes4}$ acts as $+1$ on the range of $\Pi_F$. It acts as $-1$ on the wedge, which is therefore orthogonal to that range.
\end{proof}

\begin{lemma}[Graded projector decomposition]
\label{lem:s3b-projector}
The antisymmetric Pauli projector $\Pi_F$ admits
the following decomposition.

\begin{enumerate}

\item[(i)] \textit{Full projector.}
For each two-dimensional subspace
$L\subseteq\mathbb F_2^n$, define
\[
\psi_L=\sqrt{\mathsf q}\sum_{F\in\mathbb F_2^n/L}e_F,
\]
where $e_F$ is the sorted wedge associated with the
four-element coset $F$. These vectors are orthonormal,
and
\[
\Pi_F=\sum_{\dim L=2}
|\psi_L\rangle\langle\psi_L|.
\]

\item[(ii)] \textit{Pure parity bidegrees.}
If $L\subseteq E$, every coset of $L$ lies entirely
in $E$ or $O$. Define
\[
\psi_L^\eta
=\sqrt{\frac4D}
\sum_{\substack{F\in\mathbb F_2^n/L\\F\subseteq\eta}}
e_F,
\qquad \eta\in\{E,O\}.
\]
Then
\[
\psi_L=2^{-1/2}(\psi_L^E+\psi_L^O).
\]

The compression of $\Pi_F$ to
$\Lambda^4\mathcal H_+$ is $\frac12\Pi_D$, where
\[
\Pi_D=
\sum_{\substack{L\subseteq E\\\dim L=2}}
|\psi_L^E\rangle\langle\psi_L^E|.
\]
This is the corresponding projector for $n-1$
modes, under the identification
$E\cong\mathbb F_2^{n-1}$ obtained by dropping
the last bit. The same statement holds for
$\Lambda^4\mathcal H_-$, using $O$ in place of $E$.

\item[(iii)] \textit{Mixed parity bidegree.}
If $L\not\subseteq E$, each coset contains two
even and two odd strings. Its contribution lies
in the $(2,2)$ bidegree, where
\begin{equation}
\Pi_F^{\rm mix}
=\sum_{i=1}^{\binom D2}
|f_i\wedge g_i\rangle
\langle f_i\wedge g_i|.
\label{eq:s3b-mixedproj}
\end{equation}
Here $\{f_i\}$ and $\{g_i\}$ are matched real
orthonormal bases of $\Lambda^2\mathcal H_+$
and $\Lambda^2\mathcal H_-$. Each basis vector
is supported on a perfect matching of $D/2$
disjoint edges, with coefficients
$\pm\sqrt{\mathsf q}$.

Explicitly, fix $a_0\in O$. For
$h\in E\setminus\{0\}$ and each character
$\chi$ of $E/\langle h\rangle$, the matched
basis vectors are
\begin{equation}
\begin{aligned}
f_{h,\chi}
&=\sqrt{\mathsf q}
\sum_{[x]\in E/\langle h\rangle}
\chi([x])\,e_{\{x,x\oplus h\}},\\
g_{h,\chi}
&=\sqrt{\mathsf q}
\sum_{[y]\in O/\langle h\rangle}
\chi([y\oplus a_0])\,e_{\{y,y\oplus h\}}.
\end{aligned}
\label{eq:s3b-walsh}
\end{equation}

\end{enumerate}
\end{lemma}

\begin{proof}
\emph{(i) Projector.}
In $\Pi_W$, the $Z_s^{\otimes4}$ require
$x_1\oplus\cdots\oplus x_4=0$; thus a surviving four-set
is a coset $F$ of a two-dimensional subspace $L$.
The operators $X_t^{\otimes4}$ translate these cosets.
They preserve the orientation of the sorted four-wedge:
for each nonzero difference in an affine plane, the two
corresponding unordered pairs have the same inversion
parity under translation. The number of inversions is
therefore even. Translation invariance gives the uniform
vector $\psi_L$ of Lemma S2; distinct $L$ have disjoint
supports, proving (i).

\emph{(ii) Pure parity bidegrees.}
If $L\subseteq E$, every coset has definite parity, with
$D/4$ cosets in each sector. Hence
$\psi_L=2^{-1/2}(\psi_L^E\oplus\psi_L^O)$.
Deleting the last bit identifies either parity sector,
with its ordering, with $\mathbb F_2^{n-1}$.
Thus the two diagonal pure-sector compressions of $\Pi_F$
are each $\frac12\Pi_D$.

\emph{(iii) Mixed parity bidegrees.}
When $L\not\subseteq E$, there is a unique nonzero
even direction $h\in L\cap E$, and each coset consists
of an even and an odd $h$-edge.
To determine their relative wedge orientation, let
$k$ be the highest set bit of $h$, and let $x,y$ be the
smaller endpoints of the two edges. Both have zero bit $k$.
With $\ell\ne k$ the highest set bit of $x\oplus y$, the
sorting sign is
\[
 \varepsilon_{h,L}=\begin{cases}
 +1,&\ell>k,\\
 (-1)^{1+h_\ell},&\ell<k.
 \end{cases}
\]
Indeed, for $\ell>k$ the two oriented edges occur as
separate blocks, while for $\ell<k$ counting cross-edge
inversions gives the stated parity. Moreover,
$x\oplus y$ is the representative of the odd coset
$\{a,a\oplus h\}$ with zero bit $k$.
The sign therefore depends only on $L$, not on its coset.
In the projector $|\psi_L\rangle\langle\psi_L|$ the
common sign cancels.

For fixed $h$, the mixed planes are indexed by the
$D/2$ elements of $E/\langle h\rangle$, translating a
chosen odd direction $a_0$.
Their even and odd $h$-edge bases are paired by these
translations. The Walsh transform on
$E/\langle h\rangle$ diagonalizes the pairing:
$\sum_t\chi(t)\chi'(t)=(D/2)\delta_{\chi,\chi'}$.
This gives the matched vectors $f_{h,\chi},g_{h,\chi}$
and the mixed projector decomposition stated in the lemma.
Different $h$ have disjoint edge supports; for fixed $h$
the Walsh characters are orthogonal. Since there are
$(D-1)D/2=\binom D2$ such vectors, both matched families
are orthonormal bases, as claimed.
\end{proof}

Each flat basis vector $f_i$ corresponds to a real
skew-symmetric matrix $S_i$, defined by
\[
f_i=\sum_{\substack{x<y\\x,y\in E}}
(S_i)_{xy}\,e_x\wedge e_y.
\]
Since its nonzero coefficients are
$\pm\sqrt{\mathsf q}$ on the edges of a perfect
matching, $S_i/\sqrt{\mathsf q}$ is an orthogonal
signed permutation matrix. Hence
$S_i^{\mathsf T}S_i=\mathsf q I_D$, and for unit
vectors $v,w\in\mathcal H_+$,
\begin{equation}
|\langle f_i,v\wedge w\rangle|^2
=|v^{\mathsf T}S_iw|^2
\le\mathsf q.
\label{eq:s3b-flat}
\end{equation}
The same estimate holds for the odd-sector basis
$\{g_i\}$.

\begin{lemma}[Diagonal rank-four estimate]
\label{lem:s3b-diag}
Let $H$ be positive semidefinite of rank at most
four on one parity sector. Denote its nonzero
eigenvalues, padded with zeros, by
$h_1,\ldots,h_4$. Then, for the corresponding
flat basis,
\begin{equation}
\sum_i
\langle f_i,\Lambda^2(H)f_i\rangle^2
\le
\mathsf q\left[
\sum_{a<b}(h_ah_b)^2
+6\prod_{a=1}^4h_a
\right].
\label{eq:s3b-diagbound}
\end{equation}
The same inequality holds with $g_i$ in place
of $f_i$ in the odd sector.
\end{lemma}

\begin{proof}
Let $v_1,\ldots,v_4$ be orthonormal eigenvectors of $H$ and $V=(v_1\ \cdots\ v_4)$ and let $e=\{a,b\}$ range over
the six edges of $K_4$. Set
\[
 t_{ab}=h_ah_b,\qquad
 y_i(ab)=\mathsf q^{-1}|\langle f_i,v_a\wedge v_b\rangle|^2.
\]
Then $y_i(ab)\ge0$ and
\[
 \langle f_i,\Lambda^2(H)f_i\rangle
   =\mathsf q\sum_{a<b}t_{ab}y_i(ab).
\]

\emph{Matching constraints.}
Represent $f_i$ by its real skew matrix $S_i$, so that
$S_i/\sqrt{\mathsf q}$ is an orthogonal signed-permutation matrix.
The $4\times4$ matrix $V^{\mathsf T}S_iV/\sqrt{\mathsf q}$
is a complex skew-symmetric contraction. Its row norms give
the vertex constraints, while each $3\times3$ principal
skew contraction has singular values $(\sigma,\sigma,0)$,
$\sigma\le1$, giving the triangle constraints:
\[
 \sum_{b\ne a}y_i(ab)\le1,\qquad
 \sum_{\{a,b\}\subset U}y_i(ab)\le1
 \quad (|U|=3).
\]
Together with nonnegativity these characterize the
matching polytope of $K_4$ (Edmonds' theorem~\cite{Edmonds1965}).
Thus $y_i=\sum_M\theta_{iM}\mathbf1_M$, where $M$ runs
over the empty matching, single edges, and the three
perfect matchings, with $\theta_{iM}\ge0$ and
$\sum_M\theta_{iM}=1$.

\emph{Summation over the flat basis.}
Completeness of $\{f_i\}$ implies
$\sum_i y_i(e)=\mathsf q^{-1}$ for each edge $e$.
Put $\Theta_M=\sum_i\theta_{iM}$.
Every edge $e$ belongs to a unique perfect matching
$M=\{e,e'\}$, so the aggregate weight of the singleton
$\{e\}$ is $\mathsf q^{-1}-\Theta_M$.
By Jensen's inequality,
\[
\begin{aligned}
 \sum_i\Big(\sum_e t_e y_i(e)\Big)^2
 &\le\sum_{M=\{e,e'\}\,\mathrm{perfect}}
 \Big[\Theta_M(t_e+t_{e'})^2\\[-2pt]
 &\hspace{3.8em}+(\mathsf q^{-1}-\Theta_M)(t_e^2+t_{e'}^2)\Big]\\
 &\le \mathsf q^{-1}\sum_{M=\{e,e'\}\,\mathrm{perfect}}
          (t_e+t_{e'})^2\\
 &=\mathsf q^{-1}\left[\sum_{a<b}(h_ah_b)^2
                 +6\prod_{a=1}^4h_a\right].
\end{aligned}
\]
The second inequality uses $t_e\ge0$ and
$0\le\Theta_M\le \mathsf q^{-1}$; the last identity follows from
$t_et_{e'}=\prod_a h_a$ for every perfect matching.
Multiplication by $\mathsf q^2$ proves the lemma.
The argument for the $g_i$ basis is identical.
\end{proof}

\begin{lemma}[Fourth-moment estimate]\label{lem:s3b-4th}
Let $H$ be a positive semidefinite matrix acting on one parity sector, of rank four. Let $\rho_1,\dots,\rho_4$ be its nonzero eigenvalues, let
\begin{equation}
\tau:=\prod_{a=1}^{4}\rho_a,
\qquad
T(H):=\sum_{a<b}(\rho_a\rho_b)^4,
\end{equation}
and let
\begin{equation}
X_{ij}=\langle f_i,\Lambda^2(H)f_j\rangle.
\end{equation}
Then
\begin{equation}
\sum_{i,j}|X_{ij}|^4
\le
\CHANGE{\mathsf q^2}\big[T(H)+18\tau^2\big].
\label{eq:s3b-4thbound}
\end{equation}
The same statement holds in the odd sector with $g_i$.
\end{lemma}

\begin{proof}
\emph{Step 1: two flat columns.}
Let $c,c'\in\mathbb C^{\binom D2}$ be orthonormal with
$|c_i|^2,|c'_i|^2\le \mathsf q$, and let $\alpha,\beta\ge0$.
We first establish
\begin{equation}
 \sum_i|\alpha c_i+\beta c'_i|^4
 \le \mathsf q(\alpha^4+\beta^4+6\alpha^2\beta^2).
 \label{eq:s3b-flat4}
\end{equation}
Put $\gamma=\alpha^2$, $\gamma'=\beta^2$,
$r_i=|c_i|^2$, $r'_i=|c'_i|^2$, and
$\varrho_i=\operatorname{Re}(c_i\overline{c'_i})$.
Orthogonality gives $\sum_i\varrho_i=0$, while
$|\varrho_i|\le\sqrt{r_ir'_i}$.
Define the nonnegative deficits
\[
\begin{aligned}
 \Delta_1&=\mathsf q-\sum_i r_i^2,\qquad
 \Delta_2=\mathsf q-\sum_i(r'_i)^2,\\
 \Delta_{12}&=\mathsf q-\sum_i r_ir'_i
 \ge\tfrac12(\Delta_1+\Delta_2)
 \ge\sqrt{\Delta_1\Delta_2}.
\end{aligned}
\]
With $w_i=\gamma(\mathsf q-r_i)+\gamma'(\mathsf q-r'_i)\ge0$,
weighted Cauchy--Schwarz gives
\[
 \left|\sum_iw_i\varrho_i\right|^2
 \le(\gamma\Delta_1+\gamma'\Delta_{12})
      (\gamma\Delta_{12}+\gamma'\Delta_2).
\]
Expanding the fourth powers and using $\sum_i\varrho_i=0$
therefore bounds the difference between the right- and
left-hand sides of the displayed inequality from below by
\[
 x+y+6z-4\sqrt{(x+z)(y+z)},
\]
where $x=\gamma^2\Delta_1$, $y=(\gamma')^2\Delta_2$,
and $z=\gamma\gamma'\Delta_{12}$.
Since $z^2\ge xy$, this is at least
\[
 \bigl(\sqrt{x+y+2z}-2\sqrt z\bigr)^2\ge0,
\]
proving the claim.

\emph{Step 2: four-dimensional two-forms.}
Let $V$ be four-dimensional with unit volume form $\zeta_V$.
For $t\in\Lambda^2V$ define
$t\wedge t=2\operatorname{Pf}_V(t)\zeta_V$.
The skew normal form is
$t=\alpha v_1\wedge v_2+\beta v_3\wedge v_4$
with orthonormal $v_a$ and $\alpha,\beta\ge0$.
The two columns
$c_i=\langle f_i,v_1\wedge v_2\rangle$ and
$c'_i=\langle f_i,v_3\wedge v_4\rangle$
are orthonormal and satisfy $|c_i|^2,|c'_i|^2\le \mathsf q$
by Eq.~\eqref{eq:s3b-flat}. As
$\|t\|^2=\alpha^2+\beta^2$ and
$|\operatorname{Pf}_V(t)|=\alpha\beta$,
Step 1 yields
\begin{equation}
 \sum_i|\langle f_i,t\rangle|^4
 \le \mathsf q\bigl(\|t\|^4+4|\operatorname{Pf}_V(t)|^2\bigr).
 \label{eq:s3b-form4}
\end{equation}

\emph{Step 3: wedge squares.}
For $f=f_{h,\chi}$, expand $f\wedge f$ into pairs of
distinct $h$-edges of $E$. Each such pair spans a plane
$L\subset E$ containing $h$; the wedge-orientation sign
is constant on its cosets by Lemma~\ref{lem:s3b-projector}.
For each fixed $h$, Walsh orthogonality shows that summing
over the characters contributes twice the projector
$|\psi_L^E\rangle\langle\psi_L^E|$ for each plane
$L\ni h$. Every plane contains three nonzero such
directions, so
\[
 \sum_i|f_i\wedge f_i\rangle\langle f_i\wedge f_i|
 =6\Pi_D.
\]
Let $z_i$ be the orthogonal projection of $f_i$ onto
$\Lambda^2V$. Since $f_i\wedge f_i$ and $z_i\wedge z_i$
have the same component along $\zeta_V$,
\begin{equation}
 \begin{aligned}
 \sum_i|\operatorname{Pf}_V(z_i)|^2
 &=\tfrac14\sum_i|\langle\zeta_V,f_i\wedge f_i\rangle|^2\\
 &=\tfrac32\langle\zeta_V,\Pi_D\zeta_V\rangle\le3\mathsf q.
 \end{aligned}
 \label{eq:s3b-pf-bound}
\end{equation}
The last inequality uses the four-dimensional compression
bound $\langle\zeta_V,\Pi_D\zeta_V\rangle\le4/D=2\mathsf q$.
It follows from Theorem~22 of Ref.~\cite{Sierant2026Spectral}
for $D\ge8$; for $D=4$ it is immediate.

\emph{Step 4: conclusion.}
Take $V=\operatorname{Ran}H$, put
$t_j=\Lambda^2(H)f_j\in\Lambda^2V$, and let $z_j$ be the
projection of $f_j$ onto $\Lambda^2V$.
Then $X_{ij}=\langle f_i,t_j\rangle$,
$\|t_j\|^2=(X^2)_{jj}$, and
$\operatorname{Pf}_V(t_j)=\tau\operatorname{Pf}_V(z_j)$.
Step 2, Lemma~\ref{lem:s3b-diag} applied to $H^2$,
and the bound in Step 3 yield
\[
\begin{aligned}
 \sum_{i,j}|X_{ij}|^4
 &\le q\sum_j[(X^2)_{jj}]^2
      +4q\tau^2\sum_j|\operatorname{Pf}_V(z_j)|^2\\
 &\le \mathsf q^2[T(H)+6\tau^2]+12\mathsf q^2\tau^2\\
 &=\mathsf q^2[T(H)+18\tau^2].
\end{aligned}
\]
The odd sector is identical.
\end{proof}

\begin{proof}[\CHANGE{Proof of Proposition~\ref{thm:small-gauss}}]
By Lemmas~\ref{lem:s3b-parity} and
\ref{lem:s3b-projector}, the projector $\Pi_F$
has a pure contribution on $(4,0)\oplus(0,4)$
and a mixed contribution on $(2,2)$.
Since $\Lambda^4(N)$ preserves the parity bidegrees,
$c_F(N^2)$ and $Q_{(1^4)}(N)$ split into pure and
mixed terms, which we bound separately.

\emph{Mixed part.} Define
\begin{equation}
X_{ij}=\langle f_i,\Lambda^2(A)f_j\rangle,
\qquad
Y_{ij}=\langle g_i,\Lambda^2(B)g_j\rangle.
\end{equation}
\CHANGE{On the bidegree $(2,2)$, $\Lambda^4(N)(\omega\wedge\eta)=\Lambda^2(A)\omega\wedge\Lambda^2(B)\eta$ for $\omega\in\Lambda^2\mathcal H_+$ and $\eta\in\Lambda^2\mathcal H_-$, and $\omega\otimes\eta\mapsto\omega\wedge\eta$ is an isometry. Equation~\eqref{eq:s3b-mixedproj} therefore gives}
\begin{equation}
\begin{aligned}
 c_F^{\rm mix}(N^2)&=\sum_i (X^2)_{ii}(Y^2)_{ii},\\
 Q_{(1^4)}^{\rm mix}(N)&=\sum_{i,j}|X_{ij}Y_{ij}|^2.
\end{aligned}
\end{equation}
By Lemma~\ref{lem:s3b-diag} \CHANGE{applied to $A^2$} and Lemma~\ref{lem:s3b-4th} \CHANGE{applied to $A$}, together with \eqref{eq:s3b-spectralTu},
\begin{equation}
\begin{aligned}
\sum_i[(X^2)_{ii}]^2&\le\CHANGE{\mathsf q}(T+6u^2),\\
\sum_{i,j}|X_{ij}|^4&\le\CHANGE{\mathsf q^2}(T+18u^2),
\end{aligned}
\label{eq:s3b-Xbounds}
\end{equation}
and the same estimates hold for $Y$\CHANGE{, whose invariants $u$ and $T$ are the same. The Cauchy--Schwarz inequality then gives}
\begin{equation}
 c_F^{\rm mix}(N^2)\le \CHANGE{\mathsf q}(T+6u^2),
\qquad
 Q_{(1^4)}^{\rm mix}(N)\le \CHANGE{\mathsf q^2}(T+18u^2).
\label{eq:s3b-mixedbounds}
\end{equation}

\CHANGE{\emph{Pure part.}} Since $A$ and $B$ each have rank four,
\begin{equation}
\Lambda^4(A)=u\,|\zeta_A\rangle\langle\zeta_A|,
\qquad
\Lambda^4(B)=u\,|\zeta_B\rangle\langle\zeta_B|,
\end{equation}
for unit volume forms $\zeta_A\in\Lambda^4\operatorname{Ran}(A)$ and $\zeta_B\in\Lambda^4\operatorname{Ran}(B)$\CHANGE{, and $\Lambda^4(A^2)=u^2|\zeta_A\rangle\langle\zeta_A|$, $\Lambda^4(B^2)=u^2|\zeta_B\rangle\langle\zeta_B|$. By Lemma~\ref{lem:s3b-projector}(ii) and the compression bound of
Eq.~\eqref{eq:s3b-pf-bound},
\begin{equation*}
k_A=\langle\zeta_A,\Pi_F\zeta_A\rangle=\tfrac12\langle\zeta_A,\Pi_D\zeta_A\rangle\le\mathsf q,
\end{equation*}
and likewise $k_B\le\mathsf q$. Hence
\begin{equation}
\begin{aligned}
c_F^{\rm pure}(N^2)&=u^2(k_A+k_B)\le2\mathsf q\,u^2,\\
Q_{(1^4)}^{\rm pure}(N)&=u^2\big(k_A^2+k_B^2+2|\langle\zeta_A,\Pi_F\zeta_B\rangle|^2\big)\\
&\le u^2(k_A+k_B)^2\le4\mathsf q^2u^2,
\end{aligned}
\label{eq:s3b-purebounds}
\end{equation}
where $|\langle\zeta_A,\Pi_F\zeta_B\rangle|^2\le k_Ak_B$ by the Cauchy--Schwarz inequality.}

\CHANGE{\emph{Conclusion.}} Adding \eqref{eq:s3b-mixedbounds} and \eqref{eq:s3b-purebounds}, we obtain
\begin{equation}
 c_F(N^2)\le \CHANGE{\mathsf q}(T+8u^2),
\qquad
 Q_{(1^4)}(N)\le \CHANGE{\mathsf q^2}(T+22u^2).
\label{eq:s3b-finalbounds}
\end{equation}
Substituting into \eqref{eq:s3b-sierant}\CHANGE{, with $d\mathsf q=4$,} gives
\begin{equation}
W(\phi_N)
\le
I(\mu)+168T+2688u^2.
\label{eq:s3b-precanon}
\end{equation}
It remains to identify the right-hand side with the canonical BCS value. Let $u_j=p_jq_j$. Then
\CHANGE{$u^2=u_1^2u_2^2u_3^2$, $\sum_s\lambda_s^2=\prod_j(1-2u_j)$, $\sum_s\lambda_s^4=\prod_j(1-4u_j+2u_j^2)$, and, from the pair products below Eq.~\eqref{eq:s3b-spectralTu}, $T=\sum_l(1-4u_l+2u_l^2)\prod_{k\ne l}u_k^2$. Expanding in the $u_j$ gives}
\begin{equation}
\begin{aligned}
&I(\mu)+168T+2688u^2\\
&\quad=\prod_{j=1}^{3}\big[1-4u_j+16u_j^2\big]\\
&\quad=\prod_{j=1}^{3}\big[1-4p_jq_j(1-2p_j)^2\big].
\end{aligned}
\label{eq:s3b-canonicalW}
\end{equation}
But the product on the right is precisely $2^{-\mathfrak M_2(p)}$, i.e. the stabilizer purity of the canonical BCS representative. Hence
\begin{equation}
W(\phi_N)\le 2^{-\mathfrak M_2(p)}.
\end{equation}
This is equivalent to
\begin{equation}
M_2(\phi_N)\ge \mathfrak M_2(p).
\end{equation}
Since the BCS state attains $\mathfrak M_2(p)$, the minimum over the full parity-even local unitary orbit is exactly $\mathfrak M_2(p)$. This proves the \CHANGE{proposition} for $m=3$ in the interior case. \CHANGE{For fixed $U\in G$ both sides of $W(\phi_N)\le W(\phi_M)$ are continuous in the occupations, so the bound extends to $p_j\in\{0,1\}$}, and the reductions above then give $m=2$ and $m=1$ as well.
\end{proof}

All bounds in the proof are saturated at $N=M$.
The three-mode cube supporting $M$ contains
fourteen affine planes. Two consist of the
even and odd Fock strings, each with amplitude
product $u$.

The other twelve arise from the three perfect
matchings of the four even strings.
For a matching with edge products $t,t'$
satisfying $tt'=u$, one direction gives
two planes with products $t^2,(t')^2$,
and another gives two planes with product $u$.

\subsection{Four and more pairs}
The first failure of Gaussian optimality occurs at $m = 4$ pairs. An explicit witness is a parity-even exchange of two Fock states.

\begin{proposition}[Four equal pairs]\label{prop:four}
Let $m=4$, $p_j\equiv p$, $q=1-p$, and let $\Pi$ exchange
$\ket{1000}$ and $\ket{0111}$ on each side, acting as the identity on
all other Fock states. If $\phi_s$ is the product-labeled reference
state from Sec.~\ref{app:small}, then
\begin{equation}
\begin{aligned}
&W\big((\Pi\otimes\Pi)\phi_s\big)=W_G+\Delta W,\\
&\Delta W=168(pq)^4(1-2p)^4\big[2-(1-2p)^4\big],\\
&W_G=2^{-\mathfrak M_2(p)}=\big[1-4pq(1-2p)^2\big]^4.
\end{aligned}\label{eq:four}
\end{equation}
Therefore $M_2^{f,\NL}(\Psi_4)<\mathfrak M_2(p)$ whenever
$p\notin\{0,\frac12,1\}$.
\end{proposition}

\begin{proof}
Both exchanged strings have odd parity, so $\Pi$ is an allowed
parity-even local unitary. The product-labeled state is related to the
physical BCS state by a local parity-even Clifford (Sec.~\ref{app:small}).
It therefore suffices to compute the stabilizer purity of the
permuted reference state.

Here is the finite-sum identity needed for this computation. For
real amplitudes $v_x$ of a reference state
$\phi_v=\sum_x v_x\ket{x}_A\ket{x}_B$, extended by zero outside its
support, write $W(\phi_v)=2^{-M_2(\phi_v)}$. Then
\begin{equation}
 W(\phi_v)=\sum_{s\in\mathbb F_2^n}\;
 \sum_{x_1\oplus x_2\oplus x_3\oplus x_4=0}
 \prod_{i=1}^4v_{x_i}v_{x_i\oplus s}.
 \label{eq:Wsum}
\end{equation}
Indeed, the Pauli expectation on $\phi_v$ vanishes unless both
parties have the same bit-flip string $s$. For fixed $s$, the remaining
Pauli phases give Walsh transforms of $g_s(x)=v_xv_{x\oplus s}$;
expanding their fourth powers and using Walsh orthogonality gives
Eq.~\eqref{eq:Wsum}. Equivalently,
\begin{equation*}
W(\phi_v)=2^{-n}\sum_{s,b\in\mathbb F_2^n}
 \left[\sum_x(-1)^{b\cdot x}v_xv_{x\oplus s}\right]^4.
\end{equation*}

For four identical pairs the unswapped amplitudes are
$v_x=q^{(4-|x|)/2}p^{|x|/2}$. Let $v'$ be obtained by interchanging
$v_{1000}$ and $v_{0111}$, and put $u=pq$. Substituting the 16
amplitudes into the displayed finite sum and collecting powers of
$u$ gives the exact polynomial difference
\begin{align*}
\Delta W&=W(\phi_{v'})-W(\phi_v)\\
&=168u^4-10752u^6+43008u^7-43008u^8\\
&=168u^4(1-4u)^2\big[2-(1-4u)^2\big].
\end{align*}
Since $1-4u=(1-2p)^2$ and
$W(\phi_v)=[1-4u(1-2p)^2]^4$, this is precisely
Eq.~\eqref{eq:four}. The difference is strictly positive for
$0<p<1$, $p\ne\frac12$, proving the claim.
\end{proof}

The exchange is non-Gaussian. At $p=\frac14$ it gives
$M_2=1.1173<\mathfrak M_2=1.1982$. The resulting graded reference state places
its larger Schmidt amplitudes on the Fock
labels associated with the optimal
flat-spectrum target for the $w=1$ class
discussed in Sec.~\ref{sec:m-iid} of the main text.

\begin{corollary}[From four pairs on]\label{cor:m4}
\begin{enumerate}
\item[(i)] For every $m\ge4$, Gaussian local unitaries fail to minimize
$M_2$ on an open set of BCS occupations.
\item[(ii)] If the occupations are independent with common distribution
$\nu$ with $\nu(\{0,\frac12,1\})<1$, then
\begin{equation*}
\Pr(\text{Gaussian optimality})
 \le (1-\pi_\nu)^{\lfloor m/4\rfloor},
\end{equation*}
where $\pi_\nu>0$ is the probability that some complementary
four-pair exchange $x\leftrightarrow x\oplus1111$ improves on the
Gaussian value.
\end{enumerate}
\end{corollary}

\begin{proof}
Apply the same two-sided exchange to any four selected pairs and the
identity to all remaining modes. Additivity of $M_2$ on tensor
products preserves the strict improvement in Proposition~\ref{prop:four};
continuity makes it persist in a neighborhood of four equal
nontrivial occupations, proving (i). For (ii), choose a nontrivial
$p_0$ in the support of $\nu$. Four independent draws all sufficiently
near $p_0$ have positive probability and form a successful block.
Among $\lfloor m/4\rfloor$ disjoint blocks, Gaussian optimality
requires every block to fail the improvement test. Independence gives
the stated probability bound.
\end{proof}

Figure~3(d) of the main text illustrates the
Gaussian-optimality threshold for identical BCS
pairs with $p_j=\frac14$. Local Gaussian unitaries
are optimal for $m\le3$ (Proposition~\ref{thm:small-gauss}),
whereas the graded reference states shown in the
figure achieve smaller SRE from $m=4$ onward.

\section{Particle-number-conserving local unitaries}
\label{sec:remarks}

Local particle-number conservation is more restrictive
than parity conservation. Consequently,
$D_{\min}^{f,\mathrm{NL}}\leq D_{\min}^{U(1),\mathrm{NL}}$
and
$M_2^{f,\mathrm{NL}}\leq M_2^{U(1),\mathrm{NL}}$. 
Here $M_2^{U(1),\NL}$ denotes the minimum
SRE under local unitaries conserving
$N_A$ and $N_B$. The quantity
$\Phi_S^{U(1)}$ is the maximum squared
overlap with stabilizer targets of definite
total particle number under the same local
unitaries, and
$D_{\min}^{U(1),\NL}=-\log_2\Phi_S^{U(1)}$.

\begin{proposition}[Number-conserving conversion of identical pairs]
\label{prop:u1}

Let
\[
\psi=\bigotimes_{j=1}^{m}
\left(\sqrt q\,\ket{10}
+\sqrt p\,\ket{01}\right)
\]
consist of $m$ identical number-conserving
entangled pairs. Then the optimal fidelity
under local particle-number-conserving
unitaries, with stabilizer targets of
definite total particle number, is

\begin{equation}
 \Phi_S^{U(1)}
 =\left[\max\left\{p,q,\tfrac12+\sqrt{pq}\right\}\right]^m
 =\big[\Phi_S^{\NL}(m=1)\big]^m.
 \label{eq:u1prod}
\end{equation}

The maximum is attained by products of the
one-pair stabilizer states $|10\rangle$,
$|01\rangle$, and
$(|10\rangle+|01\rangle)/\sqrt{2}$.
Hence
$D_{\min}^{U(1),\NL}
=mD_{\min}^{f,\NL}(m=1)$.

The fidelity is unchanged by filled or empty
spectator modes on $A$ and empty spectator modes
on $B$ for $p\le\tfrac12$; for $p>\tfrac12$,
exchange the two sides.

\end{proposition}
\begin{proof}
Exchange the two sides if necessary so $p\le q$ and set $t=\sqrt{p/q}$.
The $\binom mw$ amplitudes in charge sector $N_B=w$ are all
$q^{m/2}t^w$.  Let $\ket{\sigma}$ be a stabilizer target
of Schmidt rank $2^k$.
Its reduced state satisfies
$\rho_B=2^{-k}\Pi$,, whose computational-basis diagonal is uniform
on an affine binary coset $x_0+V$, with $\dim V\ge k$: this follows
from the $Z$-type part of the stabilizer expansion of $\rho_B$.
Because the target has definite particle number, the dimension of
its $N_B=w$ Schmidt sector is
$m_w=2^k\Pr_{v\in V}(|x_0\oplus v|=w)$.
Sectorwise Schmidt alignment therefore gives
\begin{equation}
 \Phi(\psi,\sigma)\le2^{-k}q^m
 \Big[\sum_w\min\{m_w,\tbinom mw\}t^w\Big]^2.
 \label{eq:u1inter}
\end{equation}
For $k\le m$, replacing the minimum by $m_w$ turns this into
$2^kq^m(\mathbb E_{v\in V}t^{|x_0\oplus v|})^2$.
Select $k$ independent coordinate functionals on $V$; their values
are independent fair bits.  Discarding the other factors (each at
most one, as $t\le1$) shows
$\Phi(\psi,\sigma)\le q^m[(1+t)^2/2]^k$,
which is at most
$[\max\{q,\frac12+\sqrt{pq}\}]^m$.
For $k>m$, instead replace the minimum in \eqref{eq:u1inter}
by $\binom mw$ and sum the binomial expansion, giving
$\Phi\le2^{-k}q^m(1+t)^{2m}
\le(\frac12+\sqrt{pq})^m$.
Product stabilizers attain the largest of the one-pair alternatives.
The same coset argument accommodates the stated spectator modes.
\end{proof}

\end{document}